\documentclass[12pt]{article}
\usepackage{graphicx}
\usepackage{latexsym,amsfonts, amssymb,epsfig}
\usepackage{amssymb,amsmath,amsthm}
\usepackage[english]{babel}

\begin{document}

\begin{center}

{\Large\bf Wave patterns within the generalized
convection--reaction--diffusion equation}\footnote{The research was
supported by the AGH local grant}

\vspace{10mm}

{\it {\large\bf V. Vladimirov}\footnote{{\it E-mail address:}
vsevolod.vladimirov@gmai.com }
\\
\vspace{5mm}

Faculty of Applied Mathematics \\
  University of Science and Technology\\
Mickiewicz Avenue 30, 30-059 Krak\'{o}w, Poland \\
 [2ex] }

\end{center}

\vspace{10mm}

{ \footnotesize {\bf Abstract.  } A set of travelling wave solutions
to a hyperbolic generalization of the convection-reaction-diffusion
is studied by the methods of local nonlinear alnalysis and numerical
simulation. Special attention is paid to displaying  appearance of
the compactly supported soloutions, shock fronts, soliton-like
solutions and peakons }

 \vspace{3mm}


\vspace{3mm}


\vspace{3mm}

\section{Introduction}

As is well-known, there do not exist methods of  obtaining the
general solutions to most of non-linear  evolutionary PDEs. Very
often in such circumstances the only alternative to numerical
studies of nonlinear models deliver the symmetry-based methods
\cite{Olv}. An important sub-class of the self-similar solutions is
formed by the travelling wave (TW) solutions. The object of this
study is to demonstrate the existence of a number of localized
self-similar TW solutions, basing on the hyperbolic generalization
of the convection-reaction-diffusion equation. We pay special
attention to the existence of solitary waves \cite{Dodd}, shock
fronts \cite{Fickett} and some other generalized solutions, such as
compactons and peakons \cite{Ros_Hyman,Li_Olver, Li_Olver_Rosenau}.
The structure of the study is following. In section 2 we introduce
our model equation and next factorize it to an ODE, describing the
set of TW solutions. Next we discuss the geometric interpretation of
solitary waves and outline the way of capturing them. In sections 3
and 4 we perform the local nonlinear analysis of the dynamical
system, equivalent to the factorized ODE, purposed at stating
conditions of the wave patterns occurrence.  In section 5, we
present the results of numerical study, revealing the presence of
all above mentioned types of TW solutions. Finally, in section 6, we
discuss the results obtained and outline the ways of further
investigations.

\section{ Statement of the problem}\label{statement}

 We consider the following evolutionary
 equation (referred to as GBE):
\begin{equation}\label{GBE1}
\alpha\,u_{tt}+u_t+u\,u_x-\varkappa\,\left(u^n\,u_x
\right)_x=(u-U_1)\,\varphi(u).
\end{equation}
Here $n,\,\varkappa,\, U_1$ are positive constants, $\alpha$ is
nonnegative. Equation (\ref{GBE1}) is a generalization of both
Burgers equation and the reaction-diffusion equation. Let us note,
that the term $\alpha\,u_{tt}$ appears when the memory effects are
taken into account \cite{Joseph,Makar_96,Makar_97,Kar}. Some
particular cases of equation (\ref{GBE1}) were studied in recent
years \cite{vladki04, vladki05,vladki06,vlamacz,Kar,Fahmy}. Owing to
these studies, the analytical description of a large variety of
travelling wave (TW) solutions is actually available.

Present investigations are mainly devoted to the qualitative and
numerical study of the family of TW solution to GBE. Our aim is to
show that under certain conditions the set of TW solutions contains
solitons, compactons, peakons and some other wave patterns. To put
it briefly, we maintain the notation traditionally used in more
specific sense. Thus, soliton is usually associated with the
exponentially localized invariant TW solution to a completely
integrable equations, possessing a number of unusual features
\cite{Dodd}. Some of these features are also inherited by the
compactons \cite{Pik_Ros_05,Pik_Ros_06}. We maintain the notion to
those solutions to (\ref{GBE1}), which manifest similar geometric
features as "true" wave patterns, known under these names.

Let us consider the set of  TW solutions
\begin{equation}\label{twans}
u(t,\,x)=U(\xi)\equiv U\left(x-V\,t\right).
\end{equation}
Inserting ansatz (\ref{twans}) into the GBE, one can obtain, after
some manipulation, the following dynamical system:
\begin{equation}\label{factors1}\begin{array}{l}
\Delta(U)\,\dot U=\Delta(U)\,W,    \\ \nonumber \\
\Delta(U)\,\dot
W=\left(U-V\right)\,W-\varkappa\,n\,U^{n-1}\,W^2-\varphi(U)\,(U-U_1)
\nonumber
\end{array}
\end{equation}
where $\Delta(U)=\varkappa\,U^n-\alpha\,V^2$. By analyzing the
factorized system (\ref{factors1}), we are going to formulate the
conditions contributing to the appearance of the soliton-like
solutions and the solutions with compact support, called {\it
compactons}, and some types of generalized TW solutions. Analysis
carried out, e.g. in \cite{Li_Olver,vsan_08} shows, that homoclinic
trajectories bi-asymptotic to saddle points correspond to both
soliton-like and compacton-like solutions. In the first case the
homoclinic loop is bi-asymptotic to a simple saddle, hence,  the
"time" which is necessary to penetrate such trajectory is infinite.
The closed loop representing the compacton is bi-asymptotic to a
topological saddle. As a result, the "time" of penetration is
finite. In fact, the compacton is is a compound generalized
solution. Its compactly supported nonzero part corresponds to the
closed loop, while the rest corresponds to the stationary point.

In order to "capture" the homoclinic trajectory among the other
solutions to the system (\ref{factors1}), we are going to state the
condition, which guarantee the stable limit cycle appearance. Choice
of such strategy is based upon the well-known fact that the growth
of the radius of the limit cycle in presence of a nearby saddle
point most often leads to the  homoclinic bifurcation. Application
of this prescription to the system (\ref{factors1}) occurs to have
some peculiarities, which are characterized below.

\begin{enumerate}
\item
The most natural parameter of the bifurcation is the wave pack
velocity $V$. Yet its change causes the movement of the line of
singular points $\Delta(U)=0$ (singular line for brevity) in the
horizontal direction. As will be shown below, the presence of the
topological saddle in most cases is due to the fact that the saddle
point belongs to the singular line. On account of this, the problem
of "capturing" the compacton-like solution becomes more complicated,
 for one must "synchronize" the moment of the homoclinic
bifurcation with the passage of the singular line through the saddle
point.
\item
 Presence of the singular line delivers an extra mechanism of the
limit cycle destruction, competing with the mechanism based upon the
homoclinic loop formation.
\item
When   the far end of the limit cycle approaches the singular line
just at the moment of the homoclinic loop formation, the latter
becomes flat and reminds triangle. Such loop corresponds to a
different type of the generalized solutions, called {\it peakons}
\cite{Li_Olver,Li_Olver_Rosenau}.
\item
The singular line is unmovable when $ \alpha = 0$. In this case,
corresponding to the parabolic-type model, the analysis becomes much
more easy, \cite{ROMP_09}.
\end{enumerate}

\section{Andronov--Hopf bifurcation in the system
(\ref{factors1})}\label{hopfbif}

The homoclinic bifurcation can occur in the system (\ref{factors1})
if  it has an extra stationary point. The function
\[\varphi(U)=\left(U-U_0 \right)^m\,\psi(U), \qquad 0 \le U_0<U_1, \]
considered throughout the remaining part of the work, assures the
required geometric configuration, providing that $\psi(U)$ does not
change sign within the segment $[U_0,\,U_1]$.

  To formulate the conditions which guarantee the limit cycle appearance in
vicinity of the stationary point $(U_1,\,0)$, let us consider the
Jacobi matrix
\[
J_1=
\left(\begin{array}{cc} 0 & \Delta(U_1) \\
{-\varphi}(U_1) & U_1-V
\end{array}
\right).
\]
In order that  $(U_1,\,0)$ be a center, the eigenvalues of $J_1$
should be pure imaginary. This is so if the conditions
\begin{eqnarray}
\mbox{Trace}\,{J_1}=U_1-V=0, \label{tr_limc} \\
\mbox{Det}\,{J_1}=\Delta(U_1)\left(U_1-U_0  \right)^m\, \psi(U_1)>0
\label{det_limc}
\end{eqnarray}
are fulfilled. The first condition immediately gives us the critical
value of the wave pack velocity $V_{cr_1}=U_1.$ The second one is
equivalent to the statement that both $\Delta(U_1)$ and $\psi(U_1)$
are nonzero and have the same signs.

The next thing we are going to do is a study of the stability of the
limit cycle. As is well known \cite{Has,GH}, this is  the real part
of the first Floquet index $\Re{C_1}$ that determines the stability
of the periodic trajectory. Depending on the sign of $\Delta(U_1)$,
there are two possibilities. If  $\Delta(U_1)>0$ when $V=V_{cr_1}$
or, in other words, the horizontal coordinate of the singular line
$\Delta(U_*)=0$, corresponding to the critical value of the
parameter $V$, satisfies the inequality
\[
U_{*}\left(V_{cr_1}\right)=\left[\frac{\alpha\,{V_{cr_1}}^2}{\varkappa}
\right]^{\frac{1}{n}}<U_1,
\]
then conditions  $\psi(U_1)>0$, and  $\Re{C_1}<0$ should be
fulfilled. In case when $U_*\left(V_{cr_1} \right)>U_1$,  these
parameters should have the opposite signs.

To obtain the expression for $\Re{C_1}$, the standard  formula
contained e.g. in \cite{GH} can be directly applied, provided that
our system is presented in the following form:
\begin{eqnarray}\label{cpf_1}
\left(\begin{array}{c}\dot z_1
\\\dot z_2 \end{array} \right)=\left(\begin{array}{lc} 0  & -\Omega \\ \Omega
& 0
\end{array} \right)\cdot \left(\begin{array}{c}z_1 \\z_2
\end{array}\right)+\left(\begin{array}{c}F(z_1,z_2)
\\G(z_1,z_2) \end{array}\right),
\end{eqnarray}
where $\Omega=\sqrt{\mu\cdot \nu},$ $\mu=|\Delta(U_1)|$,
$\nu=|\varphi(U_1)|$, $F(z_1,z_2)$ and $G(z_1,z_2)$ stand for
nonlinear terms. In this  (canonical) representation  $\Re\,C_1$ is
expressed as follows \cite{GH}:
\begin{eqnarray}\label{floqind}
16\,
\Re\,C_1=F_{111}+F_{122}+G_{112}+G_{222}+\frac{1}{\Omega}\left\{F_{12}\left(F_{11}+F_{22}\right)-
\right. \nonumber \\\left. -G_{12}\,\left(G_{11}+G_{22}\right)
-F_{11}\,G_{11}+F_{22}\,G_{22}  \right\}.
\end{eqnarray}
By $F_{ijk},\,\,F_{ij}$ we denote the coefficients of the function's
$F(z_1,z_2)$ monomials $z_i\,z_j\,z_k$, $z_i\,z_j$ correspondingly.
Similarly, indices $G_{ij}\, \, \,G_{ijk}$ denote the coefficient of
the second and third order monomials of the function $G(z_1,z_2)$.

A passage to the canonical variables $(z_1,\,z_2)$ can be attained
by the unified transformation. If the relations
(\ref{tr_limc})--(\ref{det_limc}) are satisfied, then, rewriting
(\ref{factors1}) in the coordinates $y_1=U-U_1,\,\,y_2=W$, we get
the following system:
\begin{equation}\label{auxcommon}
\Delta(U) \frac{d}{d\,\xi}\left( \begin{array}{c} y_1 \\y_2
\end{array} \right)=\left( \begin{array}{lc} 0 & \epsilon\,\mu \\ -\epsilon\,\nu &
0
\end{array} \right)\left( \begin{array}{c} y_1 \\y_2
\end{array} \right)+\left( \begin{array}{c} \Phi_1(y_1,\,y_2) \\\Phi_2(y_1,\,y_2)
\end{array} \right),
\end{equation}
where
\[
\epsilon=\left\{ \begin{array}{c} +1 \quad \mbox{if} \quad
\Delta(U_1)>0, \\
-1 \quad \mbox{if} \quad \Delta(U_1)<0,
\end{array} \right.
\]
\begin{eqnarray}
\Phi_1(y_1,\,y_2)=\varkappa\, n\, U_1^{n-2}\,y_1\,y_2
\left(U_1+\frac{n-1}{2}\,y_1 \right)+O\left(|y_i|^4  \right),
\qquad\qquad\qquad
\\
\Phi_2(y_1,\,y_2)=-\frac{1}{2}\,y_1^2
\left(2\dot\varphi(U_1)+\ddot\varphi(U_1)\,y_1 \right)-\varkappa n
U_1^{n-2} y_2^2\, \left(U_1+(n-1)y_1
\right)+y_1\,y_2+O\left(|y_i|^4\right). \nonumber
\end{eqnarray}
In order to pass to the standard representation, we apply the change
of coordinates:
\begin{eqnarray*}
\left(\begin{array}{c}z_1 \\z_2 \end{array}
\right)=\left(\begin{array}{lc}-\epsilon\sqrt{\nu} & 0 \\ 0 &
\sqrt{\mu}
\end{array} \right)\cdot \left(\begin{array}{c}y_1 \\y_2
\end{array}\right).
\end{eqnarray*}
This gives us the system (\ref{cpf_1}), with
\begin{eqnarray} F(z_1,z_2)=
\frac{\varkappa\,n\,U_1^{n-2}}
{\Omega }z_1\,z_2\,\left[\sqrt{\nu}\, U_1 -\epsilon\,\frac{n-1}{2}z_1 \right]+O(|z_i|^4),\label{FG_canon}\\
G(z_1,z_2)=\frac{\mu}{2\,\nu\Omega}z_1^2\left[\epsilon\,{\ddot
\varphi(U_1)}z_1-2\,\dot \varphi(U_1)\sqrt{\nu}
\right]-\epsilon\,\frac{\sqrt{\mu}}{\Omega}{z_1 \,z_2}-
\qquad\qquad\qquad \nonumber
\\ -\frac{\varkappa\, n\, U_1^{n-2}}{\Omega} z_2^2 \left[
U_1\sqrt{\nu}\,-\epsilon\,{(n-1)}\, z_1
\right]+O(|z_i|^4).\qquad\qquad\qquad \nonumber
\end{eqnarray}

The real part of the Floquet index is easily calculated from
(\ref{FG_canon}):
\begin{equation}\label{flindex}
\Re{ C_1}=-\frac{1}{16\,\Omega}\,G_{12}\,\left(G_{11}+G_{22}
\right)=-\epsilon\,\frac{1}{16\,\Omega^2\,|\varphi(U_1)|}
\left\{|\Delta(U_1)|\,\dot{\varphi}(U_1)+ \varkappa\, n
|\varphi(U_1)|\,U_1^{n-1} \right\}.
\end{equation}
The above can be summarized in the form of the following statement.

\vspace{3mm}

{\bf Theorem 1.} {\it If $\Delta(U_1)\cdot \psi(U_1)>0$, and
\begin{equation}\label{floqpos}
|\Delta(U_1)|\,\dot{\varphi}(U_1)+ \varkappa\, n
|\varphi(U_1)|\,U_1^{n-1}\,>\,0,
\end{equation}
then in vicinity of the critical value of the wave pack velocity
$V_{cr_1}=U_1$ a stable limit cycle appears. }

\section{Study of the stationary point
 $(U_0,0)$}\label{saddlesector}

 Rewriting  (\ref{factors1}) in the  coordinates $X=U-U_0$, $W$ we obtain the system
\begin{equation}\label{factors2}
\begin{array}{l}
\Delta(U_0+X)\,\dot X=\Delta(U_0+X)\,W,   \\
\Delta(U_0+X)\,\dot
W=\left[(U_1-U_0)-X\right]\,X^m\,\psi\left(U_0+X\right)-
\\
\qquad\qquad\qquad\qquad\qquad -\varkappa\,n\,
\left(U_0+X\right)^{n-1}\,W^2+\left(U_0-V+X\right)\,W,
\end{array}
\end{equation}
where $\Delta(U_0+X)=\varkappa \left[(U_0+X)^n-\alpha\,D^2 \right]$.
Our aim is to determine the conditions ensuring that the stationary
point $X=W=0$ is a topological saddle, or, at least, contains a
saddle sector in the right half-plane. The standard theory
\cite{AndrLeont} can be applied for this purpose. Our system can be
written down in the form
\begin{equation}\label{precan}
\frac{d}{d\,T}\left(\begin{array}{c} X \\W
\end{array}\right)=\left(\begin{array}{lc} 0 & \Delta(U_0) \\A & U_0-
V \end{array}\right)\left(\begin{array}{c} X \\W
\end{array}\right)+\mbox{nonl.\,\,terms},
\end{equation}
where $\frac{d}{d\,T}=\Delta(U_0+X)\,\frac{d}{d\,\xi}$,
\[  A= \left\{ \begin{array}{c}
\left(U_1-U_0\ \right)\,\psi(U_0), \quad\mbox{if}\quad m=1, \\
0,\qquad \quad\mbox{if}\,\,\,m\geq 2. \end{array} \right.
\]
The linearization matrix of the system (\ref{precan}) is nonsingular
if $m=1$ and the  point $U_*$ lies outside the segment
$[U_0,\,U_1]$. In this case the stationary point $(0,\,\,0)$ is a
simple saddle and the homoclinic trajectory corresponds to the
solitary wave solution. Out of this case, the Jacobi matrix has at
leas one zero eigenvalue. To study the behavior of dynamical system
in vicinity of a degenerated stationary point, we use the results
from \cite{AndrLeont}. Since we are interested in the case when the
trace of the Jacobi matrix is nonzero, the analysis prescribed in
Chapter IX of \cite{AndrLeont} is the following.
\begin{enumerate}

\item
Find the change of variables
$\left(U,\,X,\,T\right)\mapsto\left(x,\,y,\,\tau\right) $ enabling
to write down the system (\ref{precan}) in the standard form
\[
\frac{d\,x}{d\,\tau}=P_2\left(x,\,y\right),
\]
\[
\frac{d\,y}{d\,\tau}=y+Q_2\left(x,\,y \right),
\]
where $P_2\left(x,\,y\right)$, $Q_2\left(x,\,y \right)$ are
polynomials of degree 2 or higher.

\item

Solve the equation $y+Q_2\left(x,\,y \right)=0$ with respect to $y$,
presenting the result in the form of the decomposition
$y=a_1\,x^{\mu_1}+a_2\,x^{\mu_2}+...$.

\item

Find the asymptotic decomposition
\[
P_2\left(x,\,y(x)\right)=\Delta_m\,x^{m}+....
\]

\item
Depending on the values of $m$ and the sign of $\Delta_m$, select
the type of the complex stationary point, using the theorem 65 from
\cite{AndrLeont}.

\item

Return to the original variables $\left(U,\,X,\,T\right)$ and
analyze whether the geometry of the problem allows for the
homoclinic bifurcation appearance.

\end{enumerate}

So let us present the results obtained for the system
(\ref{factors2}).  First we assume, that $m>1,$ the statements of
the Andronov-Hopf theorem are fulfilled and the point $U_*$
satisfying the equation $\Delta(U_*)=0$ lies  outside the segment
$[U_0,\,U_1]$, when the parameter $V$ reaches the second bifurcation
value $V_{cr_2}$, corresponding to the homoclinic bifurcation. Here
we have three possibilities.

\begin{itemize}
\item

$U_* > U_1$ when the Andronov-Hopf bifurcation occurs (i.e.
$V=V_{cr_1}$). This inequality does not change up to the homoclinic
bifurcation, when $V=V_{cr_2}>V_{cr_1}$.

\item

The inequalities $U_0<U_*<U_1$ take place when $V=V_{cr_1}$ and the
inequality changes for $U_*<U_0$ when $V$ belongs to a small
neighborhood of $V_{cr_2}<V_{cr_1}$.

\item

$U_*<U_0$ when the Andronov-Hopf bifurcation occurs (i.e.
$V=V_{cr_1}$). This inequality does not change up to the homoclinic
bifurcation, when $V=V_{cr_2}<V_{cr_1}$.

\end{itemize}


{\bf Remark 1.} {\it Let us note that in the first case the
functions $\Delta(U)$ and $ \psi(U)$ are negative when
$U\in\,[U_0,\,U_1]$. In the third case both of the functions are
positive within the given interval. In the second case $ \psi(U_0)$
is positive, and the factor $\Delta(U_0)$ changes the sign from
negative to positive as $U_*$ becomes less than $U_0.$}

For $m>1$,  the canonical system is obtained by the formal change
$\left(X,\,W\right)\mapsto \left(x,\,y\right) $, and  passage to the
new independent variable $\tau=\left(U_0-V\right)\,T$, in  each of
the above cases. As a result of such transformation, we get the
following system:
\begin{equation}\label{Sing_outside}
\begin{array}{l}
\frac{d\,x}{d\,\tau}=\frac{\Delta(U_0+x)}{U_0-V}\,y,=P_2(x,\,y), \\
\\ \frac{d\,y}{d\,\tau}=y-\frac{1}{V-U_0}\left\{x y
-\varkappa\,n\,(U_0+x)^{n-1}\,y^2+x^m\,\left[
(U_1-U_0)-x\right]\left[\psi(U_0)+\psi'(U_0)\,x+.... \right]
\right\}= \\\\=y+Q_2(x,\,y). \end{array}
\end{equation}
Presenting $y$ in the form of series
$y=a_1\,x^{\mu_1}+a_2\,x^{\mu_2}+...$ and solving the equation
$y+Q_2(x,\,y)=0,$ we obtain
\begin{equation}\label{Q2}y=a_m\,x^m+....=\frac{U_1-U_0}{V-U_0}\,\psi(U_0)\,x^m+...\end{equation}
Inserting the function $y(x)$ into the RHS of the first equation, we
get
\begin{equation}\label{P2}
P_2(x,\,y(x))=-\frac{\Delta(U_0)}{(U_0-V)^2}\,(U_1-U_0)\,\psi(U_0)\,x^m+....=\Delta_m\,x^m+....
\end{equation}
Fulfillment of the statements of the Andronov-Hopf theorem implies
that $\Delta(U_1)\,\psi(U_1)>0$ when $V=V_{cr_1}.$ Previously we
assumed that function $\psi(U)$ does not change sign within the
segment $[U_0,\,U_1]$. And this is suffice to conclude that the
product remains positive, when the parameter $V$ attains the value
$V_{cr_2}$, corresponding to the homoclinic bifurcation. It is quite
evident for the cases one and three, because the line $\Delta(U)=0$
remains on the same side of the segment $[U_0,\,U_1]$. In the case 2
the situation is somewhat different, because $\Delta(U_0)$ is
negative for $V=V_{cr_1}$, while the function $\psi(U_0)$ is
positive and remains so when the parameter $V$ changes. But the
singular line $\Delta(U)=0$ is located  to the left from the point
$(U_0,\,0)$ when $V$ becomes close to $V_{cr_2}$, and then, in
accordance with the Remark 1, the product $\Delta(U_0)\,\psi(U_0)$
is positive,  when the homoclinic bifurcation occurs. So the
coefficient $\Delta_m$ in the decomposition (\ref{P2}) is always
negative. Basing on the classification given in Ch.~IX
 of \cite{AndrLeont},  it is possible to formulate the following
statement.

\vspace{3mm}

 {\bf Proposition 1. } {\it Let the statements of the Theorem~~1 be
fulfilled and the singular line $\Delta(U)=0$ lies outside the
segment $[U_0,\,U_1]$ of the horizontal axis. Then,
 for $m\,\ge\,2$ the  origin of the
system (\ref{Sing_outside}) is a topological saddle, having a pair
of outgoing separatrices tangent to the vertical axis and the pair
of incoming ones tangent to the horizontal axis, when $m=2\,k,\quad
k=1,2,...$. For $m=2\,k+1,\quad k=1,2,3...$, the stationary point is
a saddle-node with two saddle sectors lying in the right half-plane.
Two outgoing separatrices of the saddle sector are tangent to the
vertical axis while the incoming one is tangent to the horizontal
axis.}

\vspace{3mm}

In the following, we present the analysis of system's
(\ref{factors2}) behavior in vicinity of the stationary point
$(0,\,0)$, assuming that $\Delta(U_0)=0$ when $V=V_{cr_2}$. The
results of the study occur to depend on whether or not $U_0$ is
equal to zero. But both of these cases can be analyzed
simultaneously. We begin with the case $m>1$, for which  the
canonical system is obtained by the formal change
$\left(X,\,W\right)\mapsto \left(x,\,y\right) $ and  passage to the
new independent variable $\tau=(U_0-V)\,T$. As a result, we get the
following system:
\begin{equation}\label{ALaux1}\begin{array}{l}
\frac{d\,x}{d\,\tau}=\frac{\varkappa}{U_0-V}\,\sum_{k=1}^n{\frac{n!}{k!(n-k)!}U_0^{n-k}\,x^k}\,y,=P_2(x,\,y),
\\ \\\frac{d\,y}{d\,\tau}=y-\frac{1}{V-U_0}\left\{x y
-\varkappa\,n\,\left(x+U_0\right)^{n-1}\,y^2+x^m\,\left[
x-(U_1+U_0)\right]\left[\psi(U_0)+\psi'(U_0)\,x+...\right]
\right\}=\\ \\ =y+Q_2(x,\,y). \end{array}
\end{equation}
Presenting $y$ in the form of series
$y=a_1\,x^{\mu_1}+a_2\,x^{\mu_2}+...$ and solving the equation
$y+Q_2(x,\,y)=0,$ we  convince that the first term of the asymptotic
decomposition $y(x)$ coincides with (\ref{Q2}). This is not
surprising, since the second equations of the  systems
(\ref{Sing_outside}) and (\ref{ALaux1}) are identical. Inserting the
function (\ref{Q2}) into the RHS of the first equation of system
(\ref{ALaux1}), we get
\[
P_2(x,\,y(x))=\left\{\begin{array}{l}
-n\,\varkappa\,U_0^{n-1}\psi(U_0)
\frac{U_1-U_0}{\left(U_0-V\right)^2}\,x^{m+1}+....,\quad\mbox{if}\quad
 U_0>0, \\ \\ -\varkappa\,\psi(0)
\frac{U_1}{V^2}\,x^{n+m}+....,\quad\mbox{if}\quad
 U_0=0. \end{array}\right.
\]

In the case $m=1$ a passage to the canonical system is attained by
means of the transformation
\[
x=X, \quad y=W+B\,X, \quad \tau=(U_0-V)\,T,
\]
where
\begin{equation}\label{formforA}
B=\psi(U_0)\,\frac{U_1-U_0}{U_0-V}.\end{equation} In the new
variables, our system reads as follows:
\begin{equation}\label{ALaux2}\begin{array}{l}
\frac{d\,x}{d\,\tau}=\frac{\varkappa}{U_0-V}\,\left(y-B\,x
\right)\,\sum_{k=1}^n{\frac{n!}{k!(n-k)}U_0^{n-k}\,x^k}\,
,=P_2(x,\,y),
\qquad\qquad\qquad\qquad\\ \\
\frac{d\,y}{d\,\tau}=y+\frac{1}{V-U_0}\left[\varkappa
B^2\,U_0^{n-1}(n+1)+B+\psi(u_0)\right]\,x^2+....
=y+Q_2(x,\,y).\end{array}
\end{equation}
Solving  equation $y+Q_2(x,\,y)=0,$ we obtain
\[y=\frac{1}{U_0-V}\left[\varkappa
B^2\,U_0^{n-1}(n+1)+B+\psi(u_0)\right]\,x^2+....\] Inserting
function $y(x)$ into the RHS of the first equation, we finally get
\[ P_2(x,\,y(x))=\left\{\begin{array}{l}
-\varkappa\,U_0^{n-1}\psi(U_0)\,\frac{U_1-U_0}{(U_0-V)^2}\,x^2+...\quad\mbox{if}\quad
 U_0>0, \\ \\
 -\varkappa\,\psi(0)\,\frac{U_1}{V^2}x^{n+1}+...\quad\mbox{if}\quad
 U_0=0. \end{array}\right.
\]

\begin{figure}
\begin{center}
\includegraphics[width=3. in, height=2.25 in]{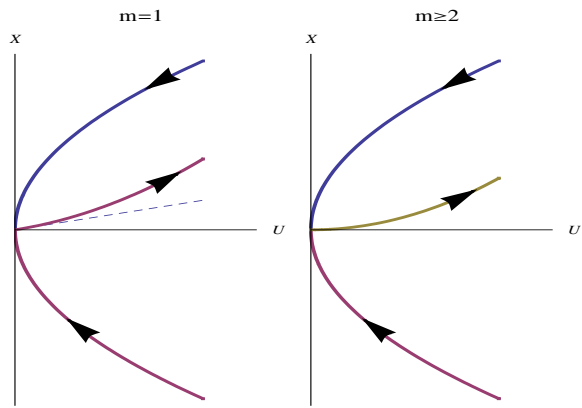}
\caption{Phase portraits of the system (\ref{factors2}) in vicinity
of the origin for different values of the parameter $m$, in cases
when $U_*\left(V_{cr_2} \right)=U_0$ }\label{fig:topsaddle1A}
\end{center}
\end{figure}

\begin{figure}
\begin{center}
\includegraphics[width=3. in, height=2.25 in]{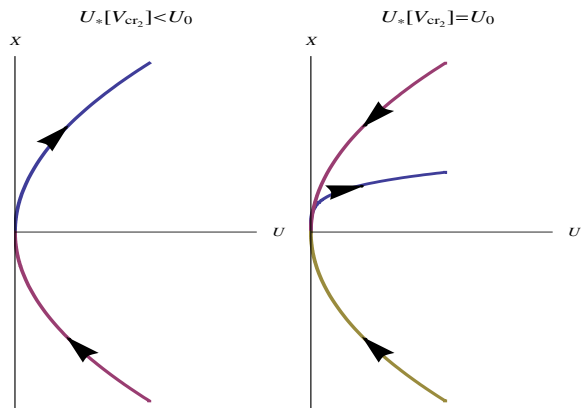}
\caption{Phase portraits of the system (\ref{factors2}) in vicinity
of the origin for $m=1/2$}\label{fig:topsaddle1B}
\end{center}
\end{figure}

So the coefficients of the lowest monomials of the decomposition of
$P_2\left(x,\,y(x)\right)$ are negative  and the following
statement, based on the classification given in \cite{AndrLeont},
holds true for $U_0=0$.

\vspace{3mm}

{\bf Proposition 2.}{\it
 \begin{enumerate} \item  If $m\geq\,2$, and  $m+n$ is an odd
natural number, then the origin of the system (\ref{ALaux1}) is a
topological saddle having a pair of outgoing separatrices tangent to
the vertical axis and the pair of incoming separatrices tangent to
the horizontal axis. For even $m+n$, the stationary point is a
saddle-node with two saddle sectors lying in the right half-plane.
Its outgoing separatrices are tangent to the vertical axis while the
incoming one is tangent to the horizontal axis. The nodal sector
lying in the left half-plane is unstable.
\item  If  $m=1$, and $n$ is an odd number then the origin of the system
corresponding to (\ref{ALaux2}) is a topological saddle identical
with that of the previous case. For even $n$, the stationary point
is a saddle-node identical with that of the previous case.
\end{enumerate}}

\vspace{3mm}

Below we formulate the analogous result for  $U_0>0$.

\vspace{3mm}

{\bf Proposition 3.}{\it
   If  $m=2\,k,\,\,k=1,2,...$, then, the origin
 of the system (\ref{ALaux1}) is a topological
saddle having a pair of  outgoing separatrices, tangent to the
vertical axis and the pair of incoming ones, tangent to the
horizontal axis. For $m=2\,k+1,\,\,k=1,2,...$, the stationary point
is a saddle-node with two saddle sectors lying in the right
half-plane. Its outgoing separatrices are tangent to the vertical
axis while the incoming one is tangent to the horizontal axis. The
nodal sector lying in the left half-plane is stable.  For $m=1$, and
arbitrary $n\in\,N$, the origin of the system corresponding to
(\ref{ALaux2}) is a saddle-node identical with that of the case
$m>1$. }

\vspace{3mm}

The last case we are going to analyze is that with $m=\frac{1}{2}$.
The motivation for such a choice will be clear later on. In order to
be able to apply the analytical theory, we rewrite the system
(\ref{factors2}), introducing new variable $Z=\sqrt{X}$:
\begin{equation}\label{ALaux}
\begin{array}{l}
\frac{d\,Z}{d\,T}=\frac{\varkappa}{2}\left[\Delta(U_0)+n\,U_0^{n-1}\,Z+...+
Z^{2\,n-1}\right]\,W, \\
\\
\frac{d\,W}{d\,T}=W\left[(U_0-V)+Z^2\right]-\varkappa\,n\,\left[U_0+Z^{2}\right]^{n-1}\,W^2+Z\,
\left[(U_1-U_0)-Z^2\right]\,\psi(U_0+Z^2).
\end{array}
\end{equation}
Let us consider the Jacobi matrix of the system (\ref{ALaux}),
corresponding to the stationary point $(0,\,0)$:
\begin{equation}\label{Jacobi_05}
J=\left[\begin{array}{cc} 0  & \frac{\varkappa}{2}\Delta(U_0)
\\ \psi(U_0)(U_1-U_0)  &  (U_0-V_{cr_2}) \end{array} \right].
\end{equation}
Analysis of the matrix (\ref{Jacobi_05}) shows, that the origin is a
simple saddle when $U_*\left(V_{cr_2} \right)$ lies outside the
segment $[U_0,\,\,U_1]$. In the case when $U_*\left(V_{cr_2}
\right)=U_0$, one of the eigenvalues of the Jacobi matrix is zero,
and, in order to classify the stationary point, we use the
 results from \cite{AndrLeont}. A passage to the
canonical variables is attained by means of the transformation
\[
x=Z, \quad y=W+B\,Z, \quad \tau=(U_0-V)\,T, \qquad
B=\frac{U_1-U_0}{U_0-V}\,\psi(U_0).
\]
The system resulting from this  is as follows:
\begin{equation}\label{ALaux3}
\begin{array}{l}
\frac{d\,x}{d\,\tau}=\frac{\varkappa}{2\,(U_0-V)}\left[n\,U_0^{n-1}\,x+...+
x^{2n-1}  \right]\,(y-B\,x)=P_2(x,\,y),
 \\ \\
\frac{d\,y}{d\,\tau}=y-\frac{1}{(V-U_0)}\left\{B\,\frac{\varkappa}{2}\left[n\,U_0^{n-1}\,x+...+
x^{2n-1}  \right]\,(y-B\,x) -x^2\,(y-B\,x)- \right.\\ \\
\left.
-\varkappa\,n\,\left[U_0+x^{2}\right]^{n-1}\,-x^3\,\psi(U_0)+...
\right\}=y+Q_2(x,\,y). \end{array}
\end{equation}
Solving the equation $y+Q_2(x,\,y)$ with respect to $y$ we obtain a
series $y=a_2\,x^2+...$. An outlook of this series' coefficients
proves to be unimportant, because they do not contribute to the
lowest term of the asymptotic decomposition of the function
$P_2(x,\,y(x))$, which is as follows:
\[
P_2(x,\,y(x))=\left\{ \begin{array}{l}  -\frac{\varkappa\,n
U_0^{n-1}\,\left(U_1-U_0
\right)\psi(U_0)}{2\,(U_0-V)^2}\,x^2+....,\quad \mbox{if}\quad U_0>0
\\ \\ -\frac{\varkappa \,\psi(0)}{2\,V^2}\,x^{2n}+....,\quad
\mbox{if}\quad U_0=0.
\end{array} \right.
\]

Let us formulate the result obtained as the following statement.

\vspace{3mm}

{\bf Proposition 4.}{\it  If  $m=1/2$, then, in case, when
$U_*\left(V_{cr_2} \right)$ lies outside the segment $[U_0,\,U_1]$,
the origin of the system (\ref{ALaux3}) is a simple saddle. In case
when $U_*\left(V_{cr_2} \right)=0$, the origin
 of the system (\ref{ALaux3}) is a saddle-node, having
two saddle sectors lying in the right half-plane. Its outgoing
separatrices are  tangent to the vertical axis while the incoming
separatrice is tangent to the horizontal axis.}

\vspace{3mm}


The crucial fact appearing from this analysis is that the stationary
points $(0,\,0)$ of the canonical systems (\ref{ALaux1}),
(\ref{ALaux2}) and (\ref{ALaux3}), depending on the values of the
parameters $m,\,n$, are either  saddles or saddle-nodes  with the
saddle sectors placed at the right half-space. The return to the
original coordinates does not cause any change in the position of
the saddle sectors, but it changes the orientation of vector fields
\footnote{We assume that $U_0-V_{cr_2}<0$, otherwise the presumable
homoclinic loop will not correspond to the wave of ''compression''.
Fulfillment of this condition is verified during the numerical
simulation} and the angles, at which the outgoing separatrices leave
the stationary point. The local phase portraits corresponding to the
distinct cases are shown in Fig.~\ref{fig:topsaddle1A},
reconstructed on the basis of the analysis of the relation between
(\ref{ALaux1})--(\ref{ALaux2}) and the system (\ref{factors2}).


Returning to coordinates $(X,\,W)$ in the case when $m=1/2$, we
obtain the patterns of the phase trajectories shown in
Fig.~~\ref{fig:topsaddle1B}. The difference between the case when
$U_*\left(V_{cr_2} \right)=U_0$ and when $U_*\left(V_{cr_2} \right)$
lies outside the segment $\left[U_0,\,U_1\right]$ is not essential -
in both cases the incoming and outgoing separatrices enter the
origin tangent to the vertical axis. Yet, since in the second case
the upper sepratrice is a mirror image of the lower one, then one
can expect the compactly supported solution to be more symmetric.

Before we start to discuss the results of numerical study of the
system (\ref{factors2}), let us analyze to what type of solitary
waves will correspond the homoclinic loops, presumably appearing
 in system (\ref{factors1}). In order to analyze this issue, we
are looking for the asymptotic solution
$W(X)=\alpha_1\,X^{\mu_1}+\alpha_2\,X^{\mu_2}+...$ of the equation
\begin{equation}\label{asymptd} \begin{array}{l}
\Delta(U_0+X)\,W\,\frac{d\,W}{d\,X}=G\equiv \\ \\ \equiv
\left[(U_0-V)+X\right)\,W-\varkappa\,n\,(U_0+X)^{n-1}\,W^2-\psi(U_0+X)\,\left[(U_0-U_1)+X\right]\,X^m,
\end{array}\end{equation}
which is equivalent to the system (\ref{factors2}). Let us start
with the case $U_0=0$, for which the equation (\ref{asymptd}) can be
re-written as follows:
\begin{eqnarray}\label{asymptd2}
\varkappa\,X^{n+2\,\mu_1-1} \left(
\alpha_1+\alpha_2\,X^{\mu_2-\mu_1}+...\right)\,\left(
\mu_1\,\alpha_1+\mu_2\,\alpha_2\,X^{\mu_2-\mu_1}+...\right)= \nonumber \\
= X^{1+\mu_1}\left(\alpha_1+\alpha_2\,X^{\mu_2-\mu_1}+... \right)-
V\,X^{\mu_1}\left(\alpha_1+\alpha_2\,X^{\mu_2-\mu_1}+... \right)-
\\
 -\varkappa\,n\,X^{n+2\,\mu_1-1}\left(\alpha_1^2+2\alpha_1\alpha_2\,X^{\mu_2-\mu_2}+....\right)-
\left[X^{m+1}+U_1\right]\,X^m\,\psi(0)+....  \nonumber
\end{eqnarray}
The procedure of solving (\ref{asymptd2}) is pure algebraic: we
collect the coefficients of different powers of $X$ and equalize
them to zero. The lowest power in the RHS is either $X^{\mu_1}$  or
$X^{n+2\,\mu_1-1}$. The number ${n+2\,\mu_1-1}$ cannot be less or
equal to $\mu_1$, because it involves the inequality $0<\mu_1 \le
1-n$, which is impossible for any natural $n$. On the other hand, if
${n+2\,\mu_1-1} \le m$, then $\mu_1$ becomes an ''orphan'' and
$\alpha_1$ should be nullified.

So $X^{n+2\,\mu_1-1}$ cannot be the lowest monomial. From this, it
immediately appears that the only choice leading to a nontrivial
solution is $\mu_1=m$.

For $U_0>0$ the equation is as follows:
\begin{eqnarray}\label{asymptd3}
\varkappa\,X^{2\,\mu_1} \left(
\alpha_1+\alpha_2\,X^{\mu_2-\mu_1}+...\right)\,\left(
\mu_1\,\alpha_1+\mu_2\,\alpha_2\,X^{\mu_2-\mu_1}+...\right)= \nonumber \\
= \psi(U_0)\,X^m\left[ (U_1-U_0)-X \right]- \nonumber
\\
 -\varkappa\,n\,\left[U_0^{n-1}+...   \right]\,\left(\alpha_1^2\,X^{2\,\mu_1}+...
 \right)+\left[X+(U_0-V)   \right]  \left[\alpha_1\,X^{\mu_1}+...
 \right].
\end{eqnarray}
Using the analogous arguments as before for the present (more
simple) case, we conclude that $\mu_1=m$.

The decomposition obtained can be used to asymptotically integrate
the equation
\[
\frac{d\,X}{d\,\xi}=W=a_1\,X^m+....
\]
which immediately gives in the lowest order the expression
\begin{equation}\label{outgoing}
X=\left\{\begin{array}{l}\left[
{a_1}{(1-m)}\,(\xi-\xi_0)\right]^{\frac{1}{1-m}}, \quad \mbox{if}
\quad m\neq 1, \nonumber
\\ \\C\,\exp{\left[a_1 \,(\xi-\xi_0)\right]}, \quad\quad\quad\,\,\, \mbox{if}
\quad m = 1,\end{array} \right.
\end{equation}
from which we conclude that the trajectory reaches the origin in
''finite time'' if $m<1.$

Unfortunately, in case when $m\in N_+$, presented above result
cannot be attributed to  both of the saddle sector separatrices,
forming the closed loop. In fact, the incoming separatrix of the
stationary point $(0,\,0)$ in all cases considered here  is tangent
to the vertical axis and therefore cannot be described by the
formula (\ref{outgoing}) when $m \geq 1.$ The above formula, then,
describes the asymptotic behavior of the ''tail '' of the solitary
wave, corresponding to the homoclinic solution.

To complete the analysis, we resort to some arguments, concerning
the other end of the homoclinic trajectory. It is the common feature
of almost all the cases considered here, that the incoming
separatrice is tangent to the vertical axis, with the exception of
the case when $m=1$ and $U_*\left(V_{cr_2} \right)$ lies outside the
interval $(U_0,\,U_1)$. This enables us to assume that in vicinity
of the origin $W=-B\,U^{\sigma}+o\left(U^{\sigma}\right)$, where
$0<\sigma<1$ and $B>0$. An approximate equation describing the first
coordinate of the separatrix is, then, as follows
\[
\frac{d\,U}{d\,\xi}=-B \,U^\sigma +o\left(U^{\sigma}\right).
\]
Performing the asymptotic integration, we obtain, up to the change
of notation, the solution identical with (\ref{outgoing}),
concluding from this that in all analyzed above cases with
$m,\,\,n\,\in\,N_+,$ the incoming separatix reaches the origin in
finite ''time''.

It is obvious, that similar arguments can be applied to the analysis
of presumable homoclinic trajectory, corresponding to $m=1/2$. It
appears from the above analysis, that the  incoming and outgoing
separatrces are tangent to the vertical axis and both of them reach
the origin in finite "time". Let us note that when
$U_*\left(V_{cr_2} \right)$ lies outside $[U_0,\,U_2],$  the
asymptotic behavior of the spearatrices is described by the formula
\[
X=\pm\,\sqrt{\gamma\,(\xi-\xi_0)}, \qquad \gamma>0.
\]
So this is the only case when the homoclinic trajectory corresponds
with certain to the compactly supported solution of the equation
(\ref{GBE1}).

\section{Results of numerical simulation}\label{numerical}

\begin{figure}
\begin{center}
\includegraphics[totalheight=1.5 in]{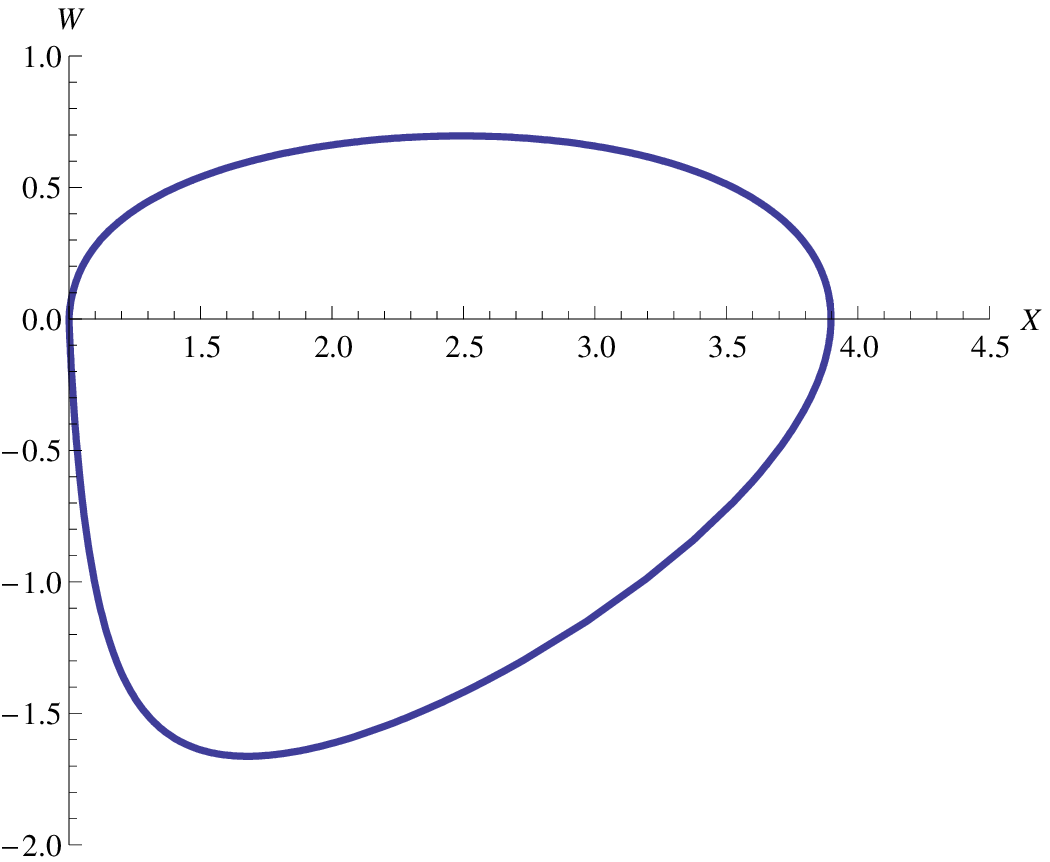}
\hspace{5 mm}
\includegraphics[totalheight=1.5 in,origin=c]{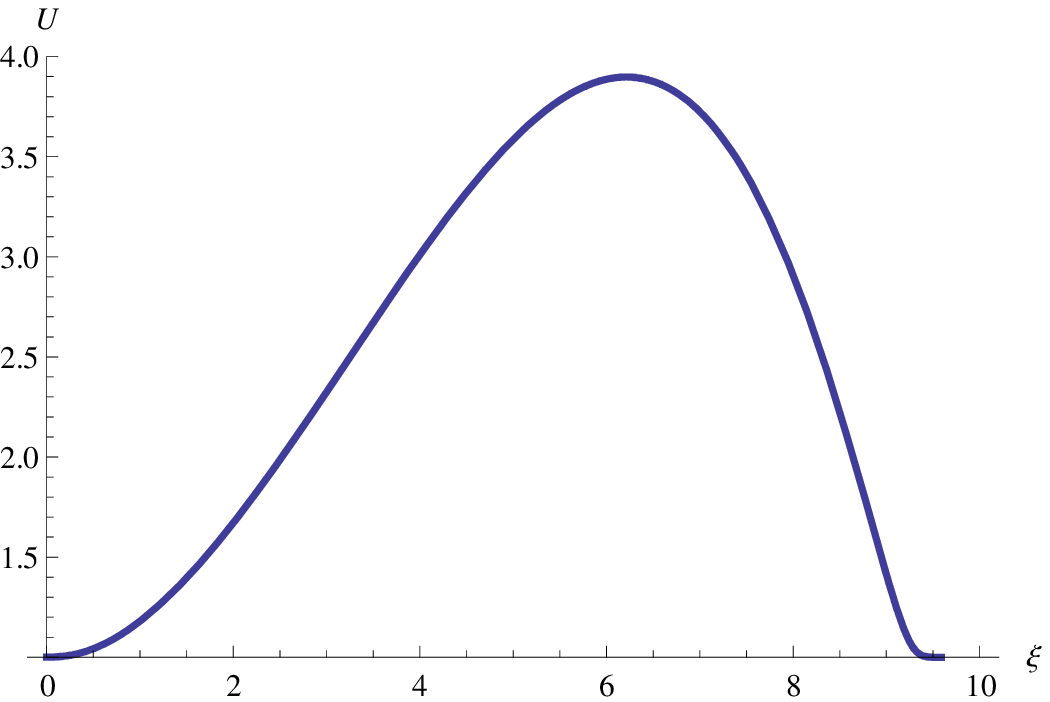}
\caption{Homoclinic solution of the system (\ref{factors2}) with
$\varphi(U_0+X)=X^{1/2}$ (left) and the corresponding compactly
supported TW solution to Eq. (\ref{GBE1}) (right), obtained for
$n=1$, $\alpha =0.12$, $V_{cr_2} \cong 2.68687$ and
$U_*-U_0=-0.133684$ }\label{fig:2}

\end{center}
\end{figure}

\begin{figure}
\begin{center}
\includegraphics[totalheight=1.5 in]{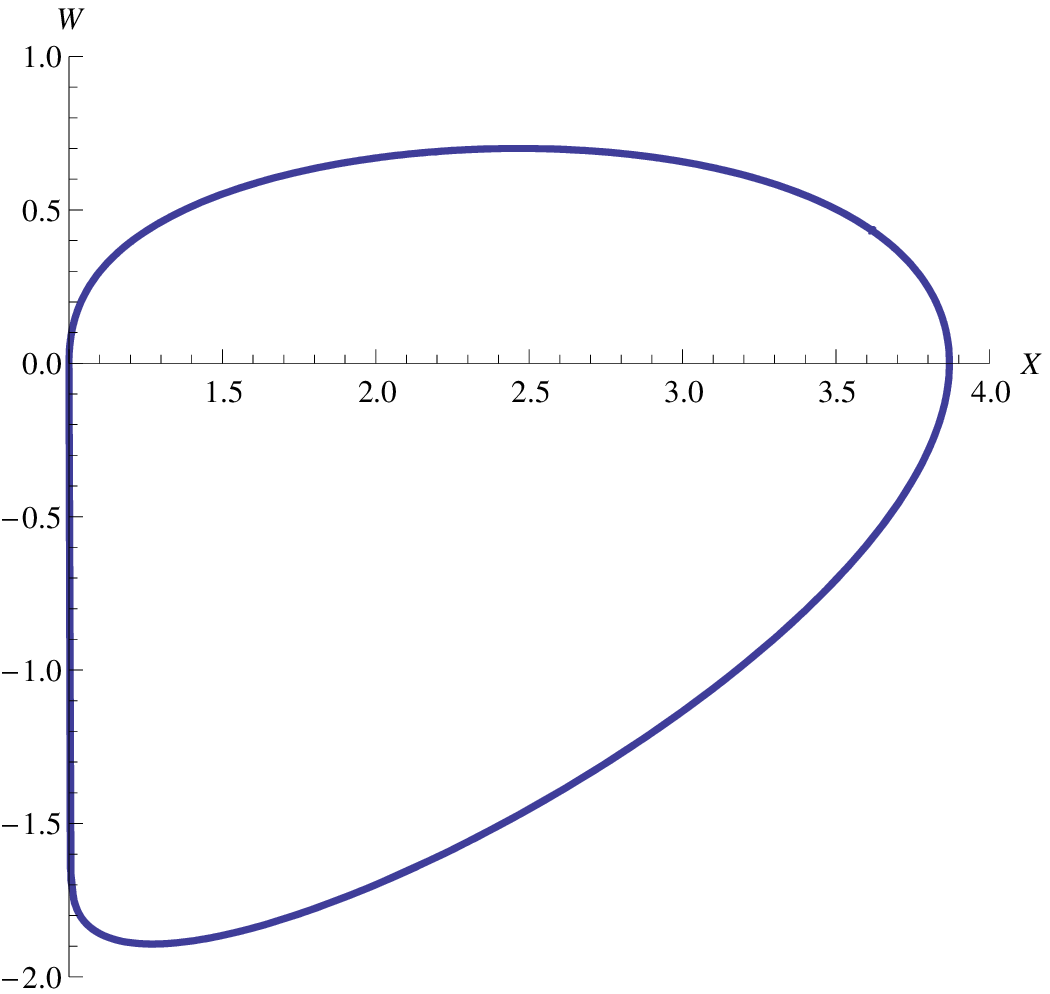}
\hspace{5 mm}
\includegraphics[totalheight=1.5 in,origin=c]{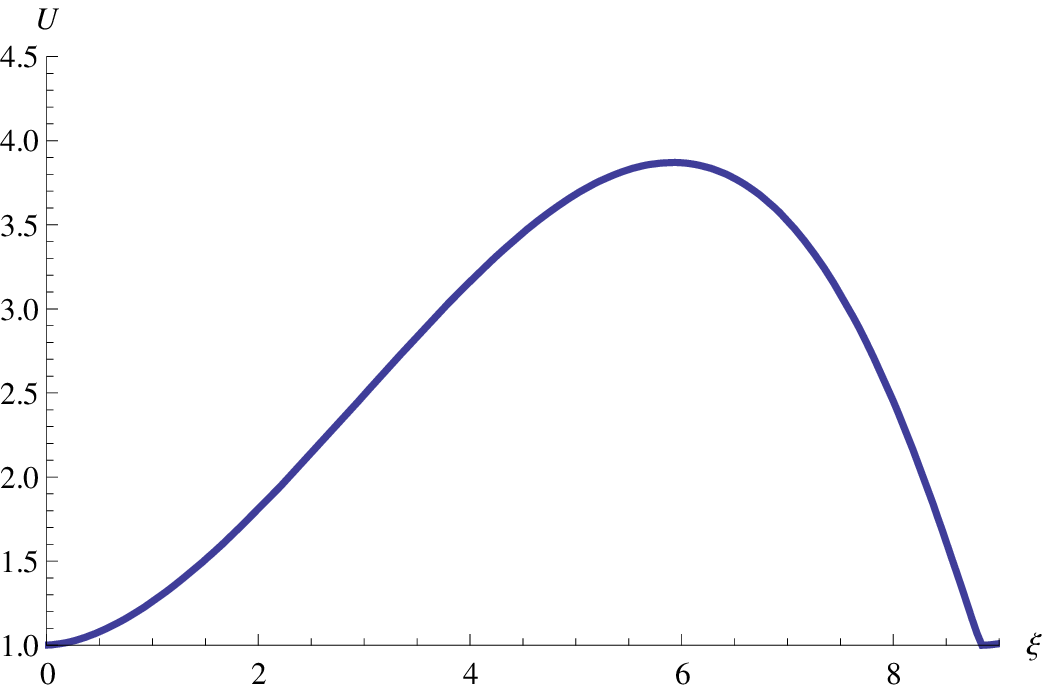}
\caption{Homoclinic solution of the system (\ref{factors2}) with
$\varphi(U_0+X)=X^{1/2}$ (left) and the corresponding TW solution to
Eq. (\ref{GBE1}) (right), obtained for $n=1$, $\alpha= 0.13827$,
$V_{cr_2} \cong 2.68892$ and $U_*-U_0=0.99973$ }\label{fig:3}

\end{center}
\end{figure}

Since the parabolic case was discussed in detail in our previous
work \cite{ROMP_09}, we mainly concentrate here on the hyperbolic
case, corresponding to $\alpha>0$. Numerical simulations of the
system (\ref{factors2}) were carried out with $\varkappa=1,\,\,
U_1=3,\,\,U_2=1$. The remaining parameters varied from one case to
another. The results of qualitative study evidence that, within the
variety of parameters that were analyzed, the only case when when
the homoclinic loop corresponds to the compactly supported solution
is that with $m=1/2$. We discuss the results concerning the details
of the phase portraits in terms of the reference frame $(X,\,W)$.
The numerical experiments show that in case when $m=1/2,\,\,n=1$ and
$U_*\left(V_{cr_2}\right)<U_0$, both of the separatrices forming
closed loop enter the stationary point $(0,\,\,0)$ tangent to the
vertical axis, Fig.~~\ref{fig:2}, the left picture. The right
picture shows  the compactly supported solution to Eq. (\ref{GBE1}),
corresponding to the closed loop. When $U_*\left(V_{cr_2}\right)$
tightly approaches $U_0$, no matter from the left of from the right,
the left side of the homoclinic trajectory is clasped to the
vertical axis, Fig.~~\ref{fig:3}, left. The corresponding compactly
supported solution has more sharp front, Fig.~~\ref{fig:3}, right.

\begin{figure}
\begin{center}
\includegraphics[totalheight=1.2 in]{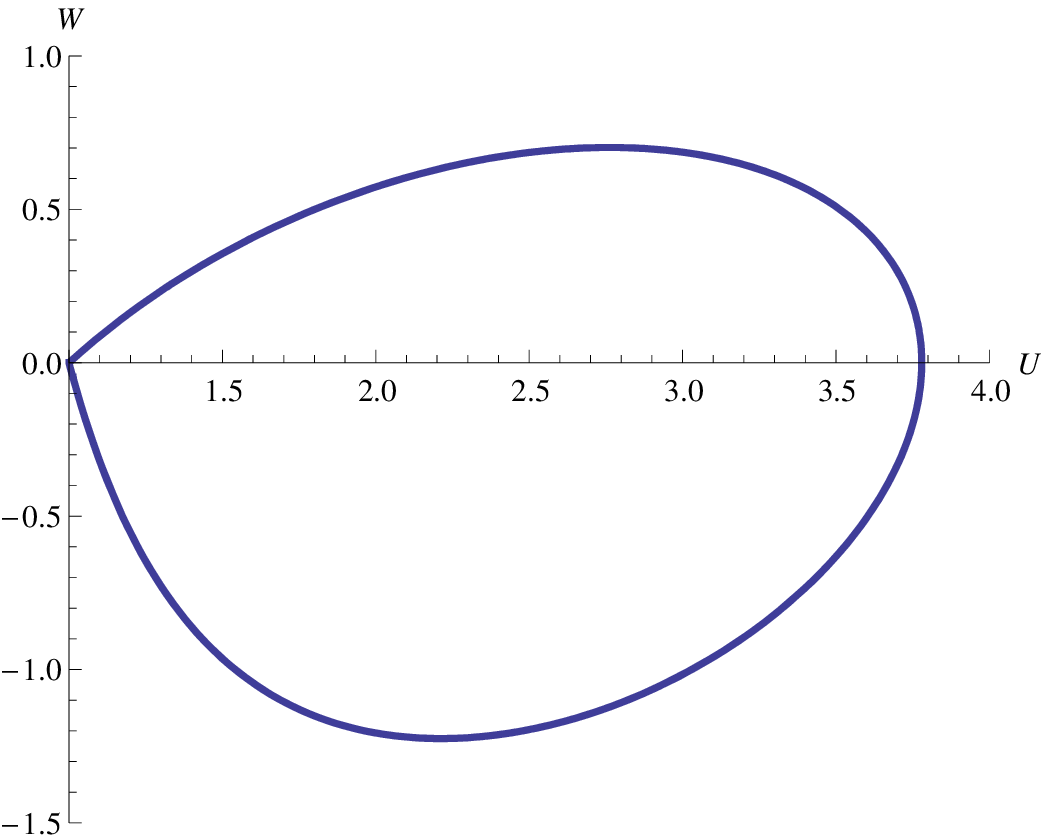}
\includegraphics[totalheight=1.2 in,origin=c]{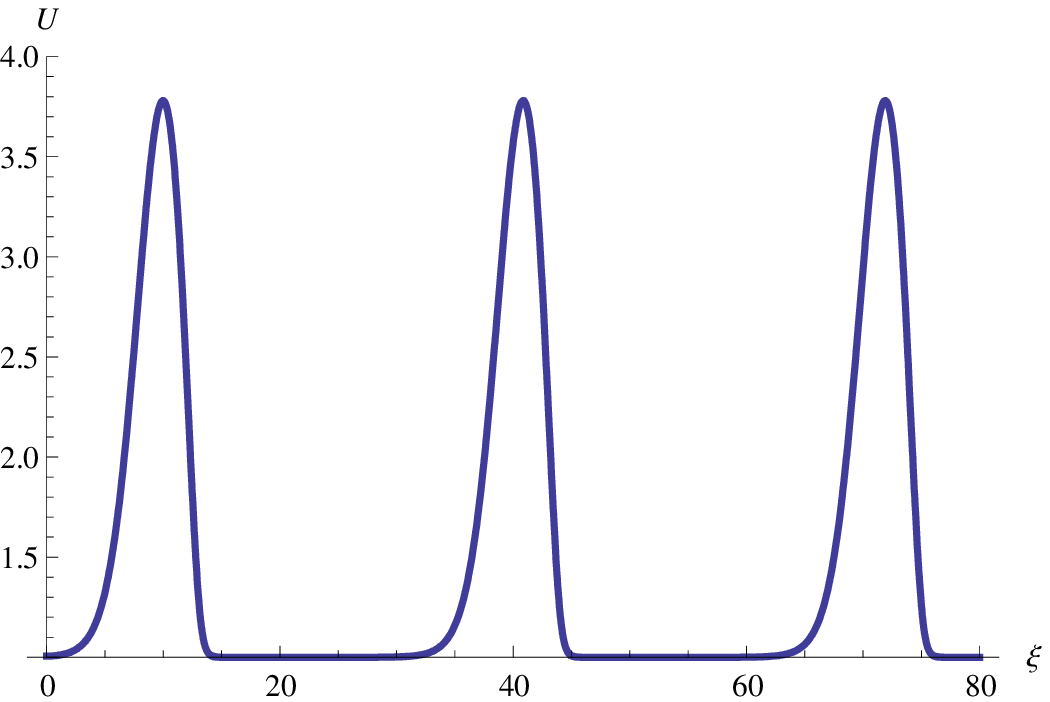}
\includegraphics[totalheight=1.2 in,origin=c]{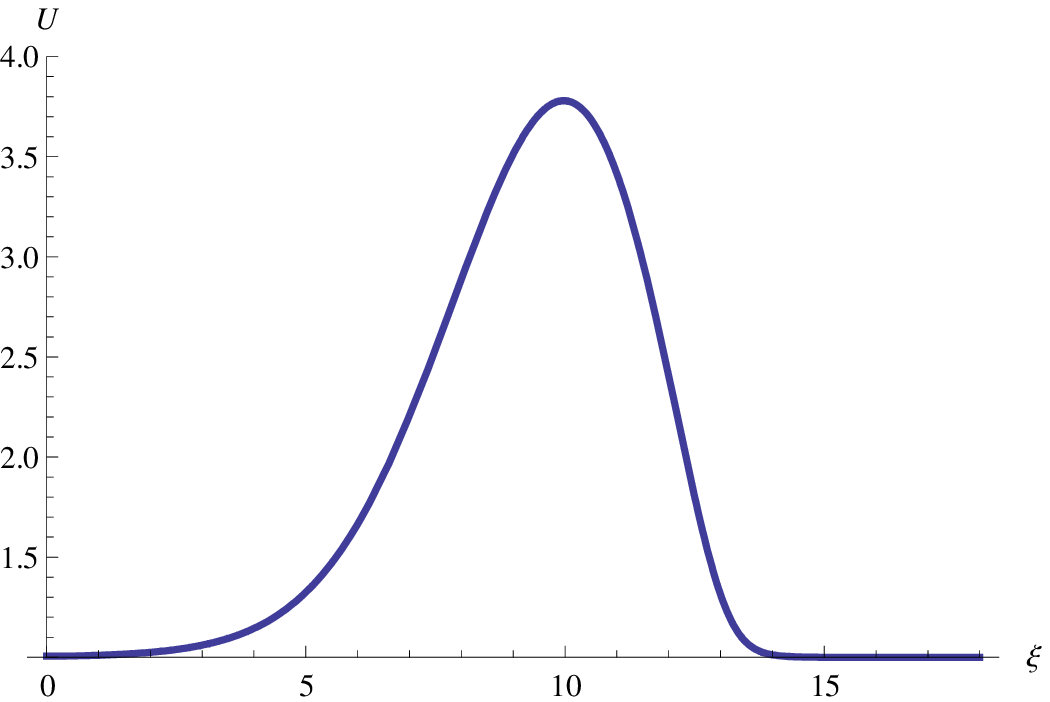}
\caption{Homoclinic solution of the system (\ref{factors2}) with
$\varphi(U_0+X)=X^1$ (left),  the corresponding tandem of
well-localized soliton-like solutions to Eq. (\ref{GBE1}) (center),
and the soliton-like solution (right), obtained for $n=1$, $\alpha=
0.06$, $V_{cr_2} \cong 2.65795$ and $U_*-U_0=-0.576119$
}\label{fig:4}

\end{center}
\end{figure}

As it was shown in the previous section, appearance of the
soliton-like solutions is possible merely in case when $m=1$. In
accordance with the predictions of our qualitative analysis, such
solutions were observed for $m=n=1$ and  values of the paremeters
guaranteeing the fulfillment of the inequality $U_*\left(V_{cr_2}
\right)<U_0$. As it is seen on the left picture of
Fig.~~\ref{fig:4}, both of the separatrices of the saddle sector
form nonzero angle with the vertical axis and this gives us the
reason to state that the homoclinic loop corresponds in this case to
the smooth solitary wave, which is nonzero for any $\xi\,\in\,R$.

\begin{figure}
\begin{center}
\includegraphics[totalheight=1.2 in]{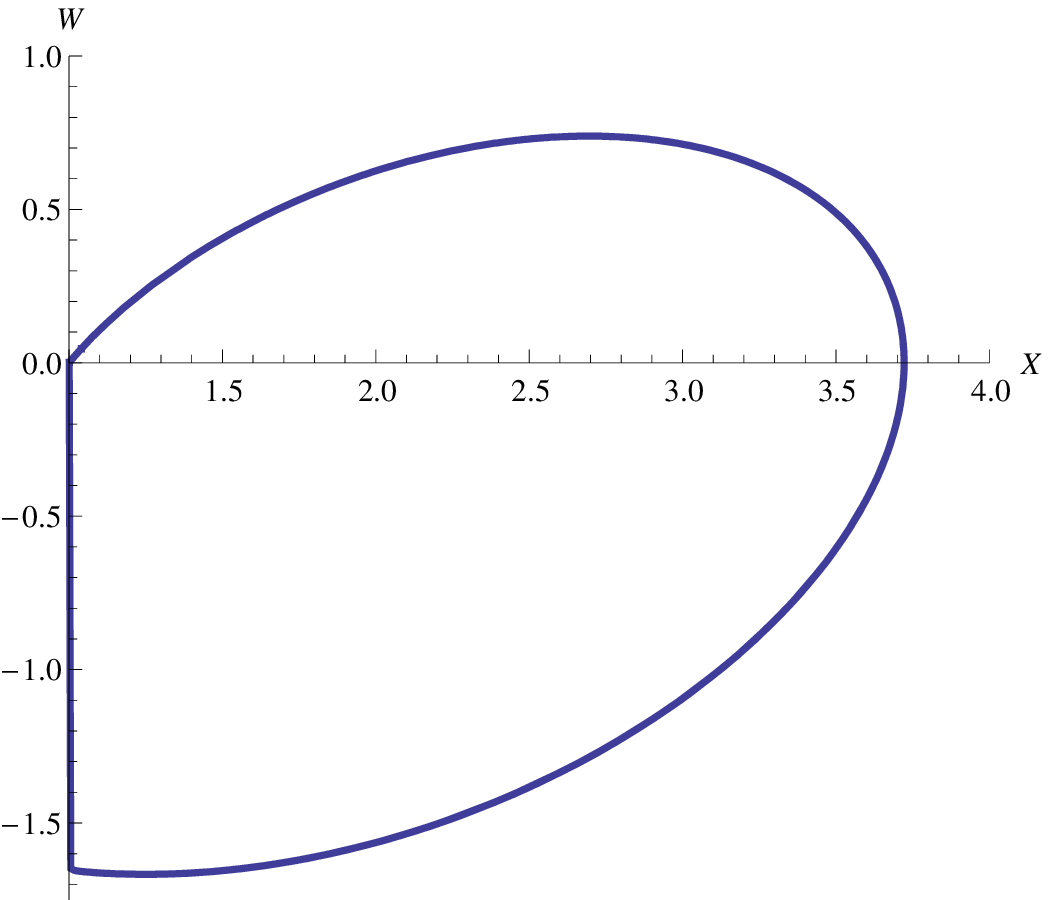}
\includegraphics[totalheight=1.2 in,origin=c]{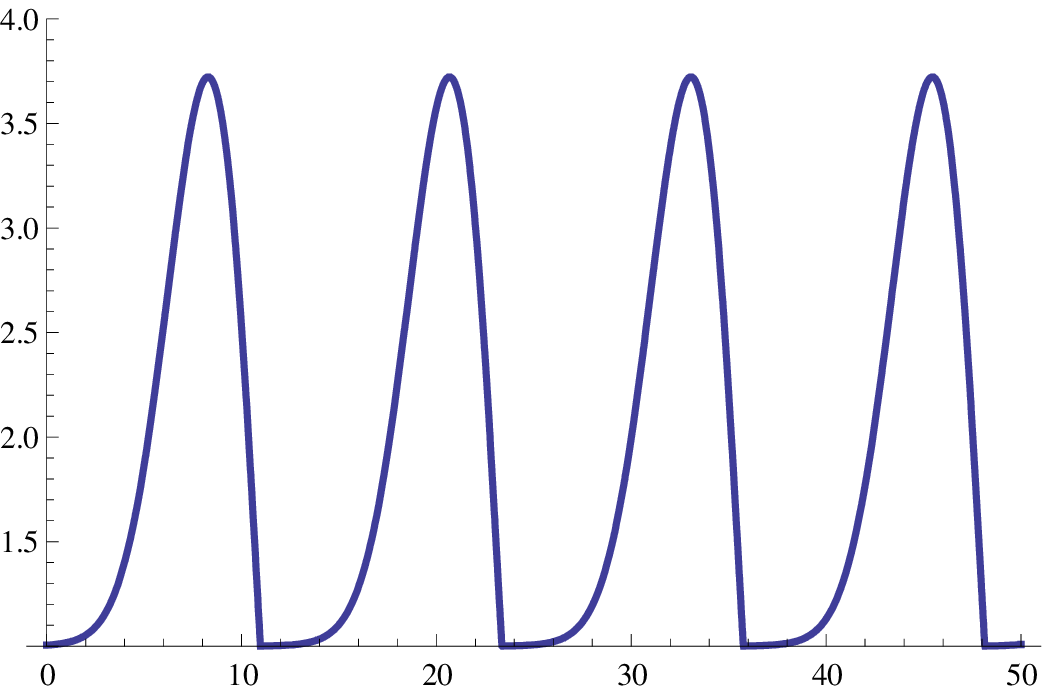}
\includegraphics[totalheight=1.2 in,origin=c]{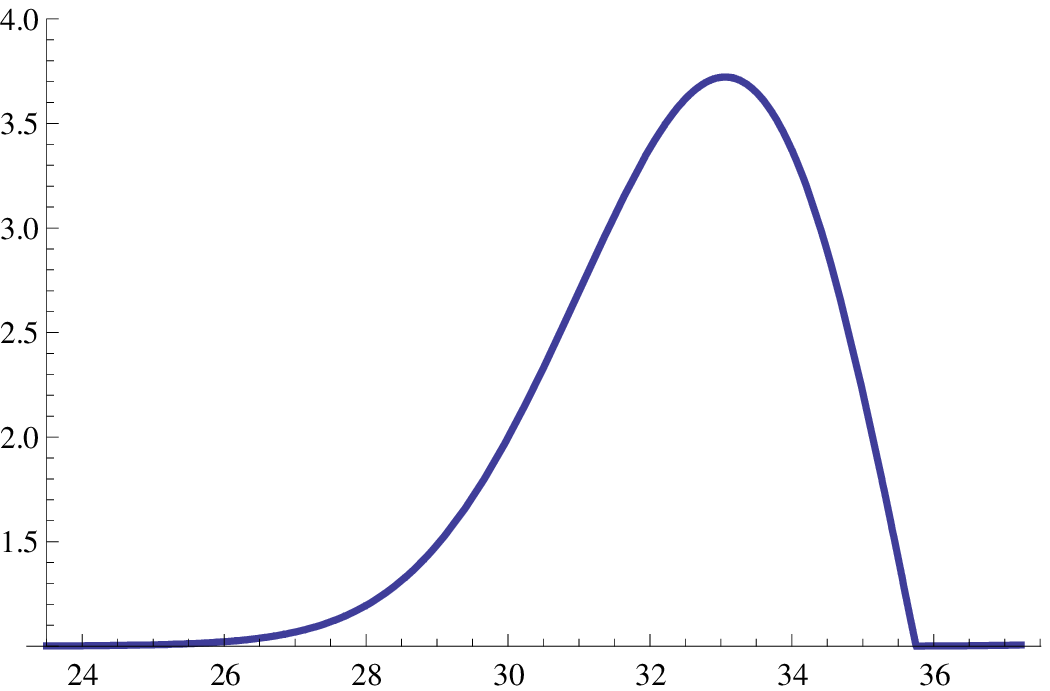}

\caption{Homoclinic solution of the system (\ref{factors2}) with
$\varphi(U_0+X)=X^1$ (left), the corresponding tandem od solitary
wave solutions to Eq. (\ref{GBE1}) (center) and a single solitary
wave solution (right), obtained for $n=1$, $\alpha= 0.142$,
$V_{cr_2} \cong 2.65489$ and $U_*-U_0=0.000878617$ }\label{fig:5}

\end{center}
\end{figure}

When $U_*$ tightly approaches $U_0$, the incoming separatrice of the
homoclinic trajectory is clasped to the vertical axis
(Fig.~~\ref{fig:5}, left) and the solutions reminding shock waves
with relaxed tails are observed in place of soliton-like wave packs
(Fig.~~\ref{fig:5}, center, right).

\begin{figure}
\begin{center}
\includegraphics[totalheight=1. in]{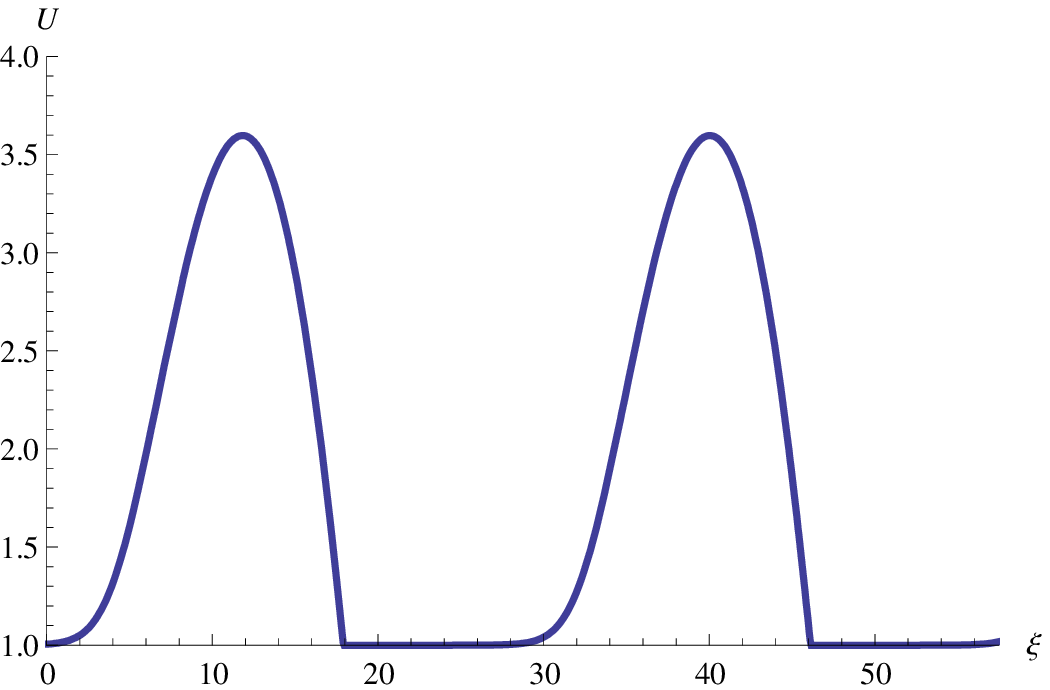}
\includegraphics[totalheight=1. in,origin=c]{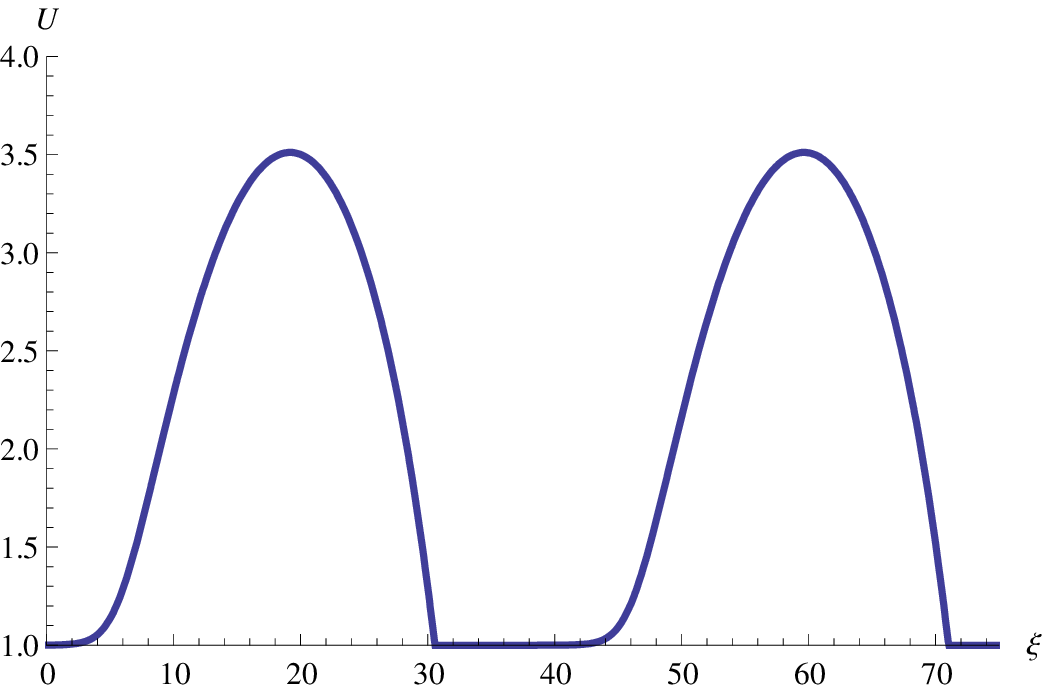}
\includegraphics[totalheight=1. in,origin=c]{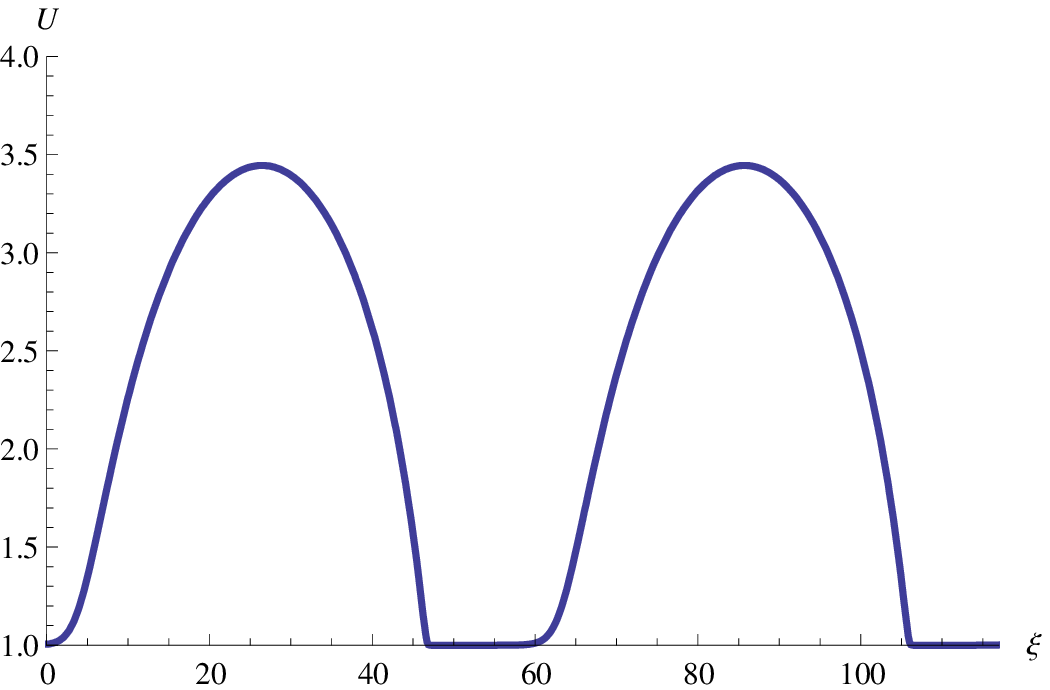}

\caption{Tandems od shock-like solutions to Eq. (\ref{GBE1}),
corresponding to $\varphi(U_0+X)=X^1$, $U_* \approx U_0$, $n=2$
(left), $n=3$ (center), and $n=4$ (right)}\label{fig:6}

\end{center}
\end{figure}

For $m\geq 2$  merely the shock-like solutions have been observed in
numerical experiments. This confirms the arguments put forward in
the previous section. The outlook of  the TW occurs to depend on the
values of the parameters $m,\,n$. A series of shock-like solutions
corresponding to $m=1$ and $n=2,\,3,\,4$ are shown in
Fig.~~\ref{fig:6}. It is seen, that effective width of the TW grows
as $n$ grows, and the shape becomes more and more gently sloping.

\begin{figure}
\begin{center}
\includegraphics[totalheight=1. in]{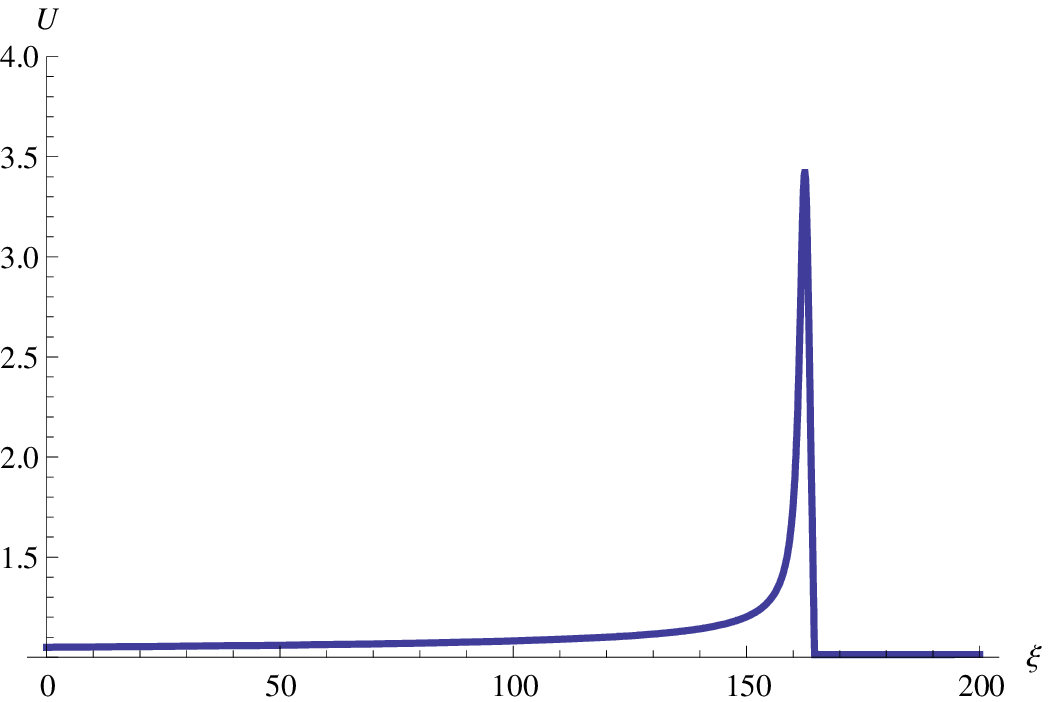}
\includegraphics[totalheight=1. in,origin=c]{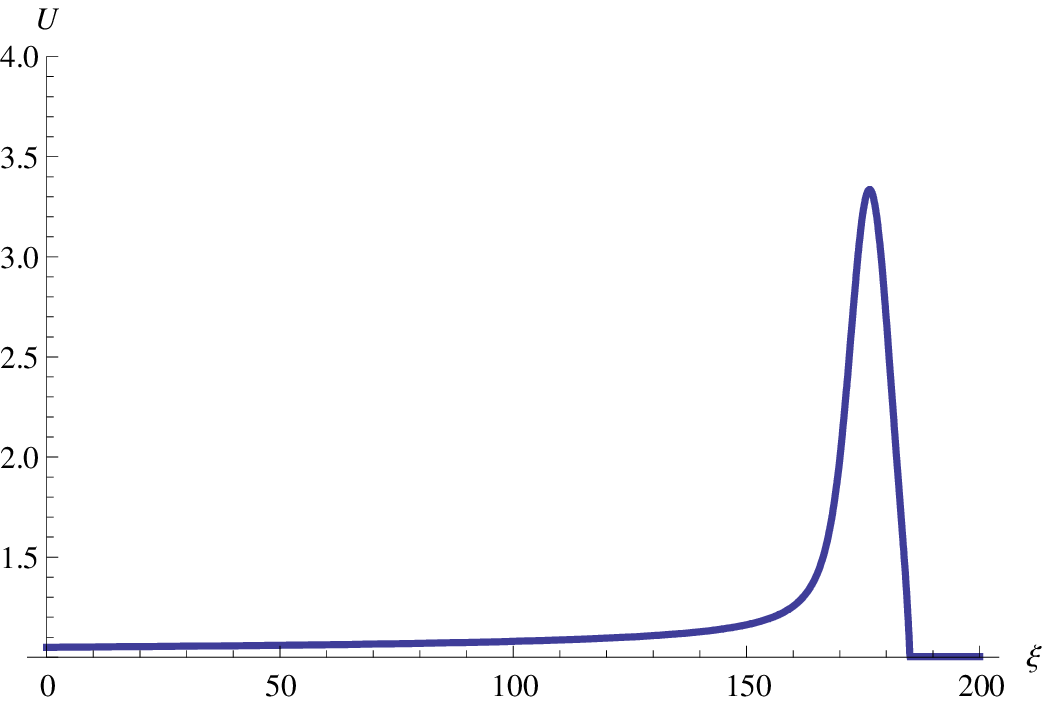}
\includegraphics[totalheight=1. in,origin=c]{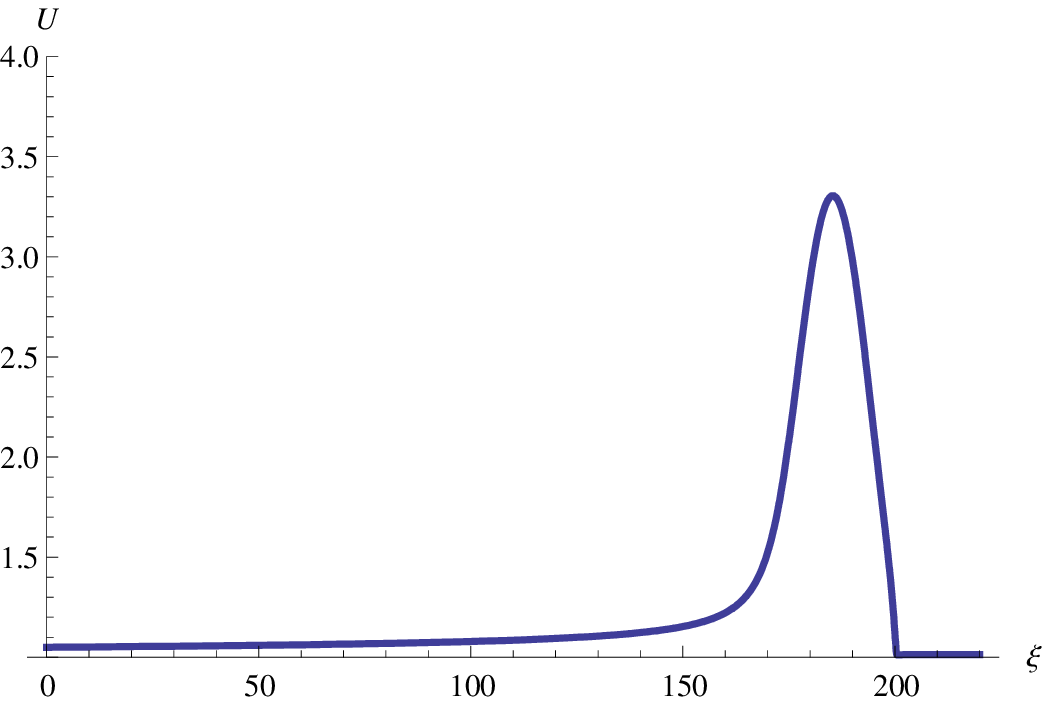}

\caption{ Shock-like solutions to Eq. (\ref{GBE1}), corresponding to
$\varphi(U_0+X)=X^3$, $U_* \approx U_0$, $n=1$ (left), $n=3$
(center), and $n=4$ (right)}\label{fig:7}

\end{center}
\end{figure}

The next series (Fig.~~\ref{fig:7}) shows the TW corresponding to
$m=3$ and $n=1,\,2,\,3$. This picture differs from the previous one
in that the "tails" of the travelling waves are longer. A common
feature of all the cases with $m\geq\,2$ is the strong stability of
the equilibrium, especially in the direction of the outgoing
separatrice. Choosing the initial data more and more close to the
origin, we are able to elongate the "tail" without limit. Besides,
the profiles of the TW  become more and more smooth as the $n$
grows.

\begin{figure}
\begin{center}
\includegraphics[totalheight=1. in]{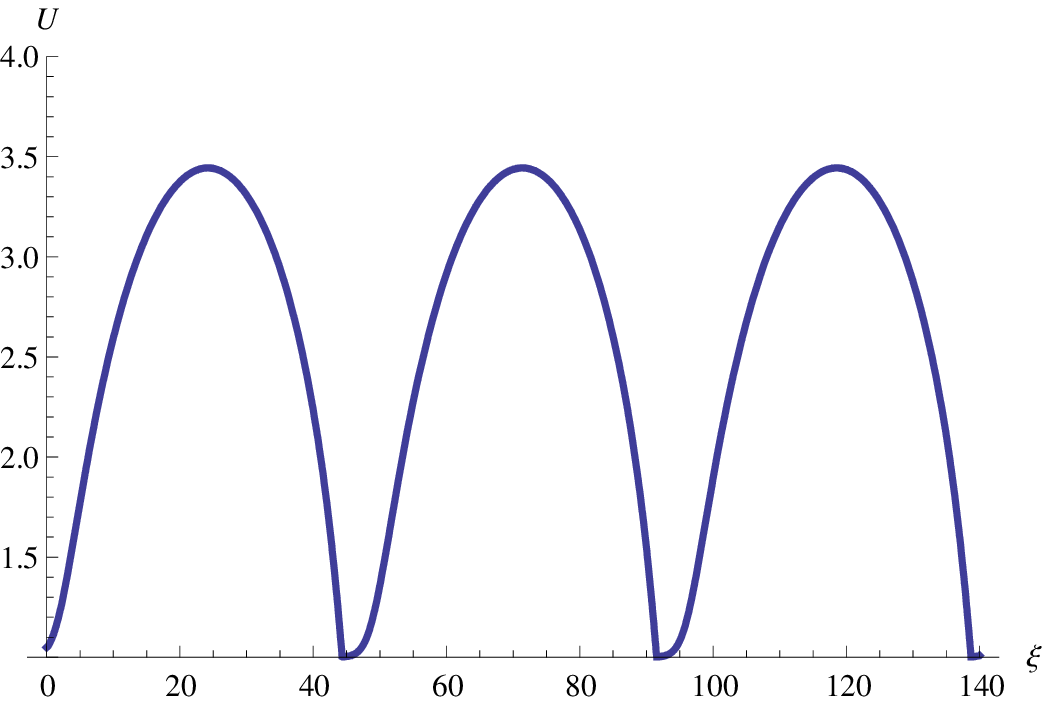}
\includegraphics[totalheight=1. in,origin=c]{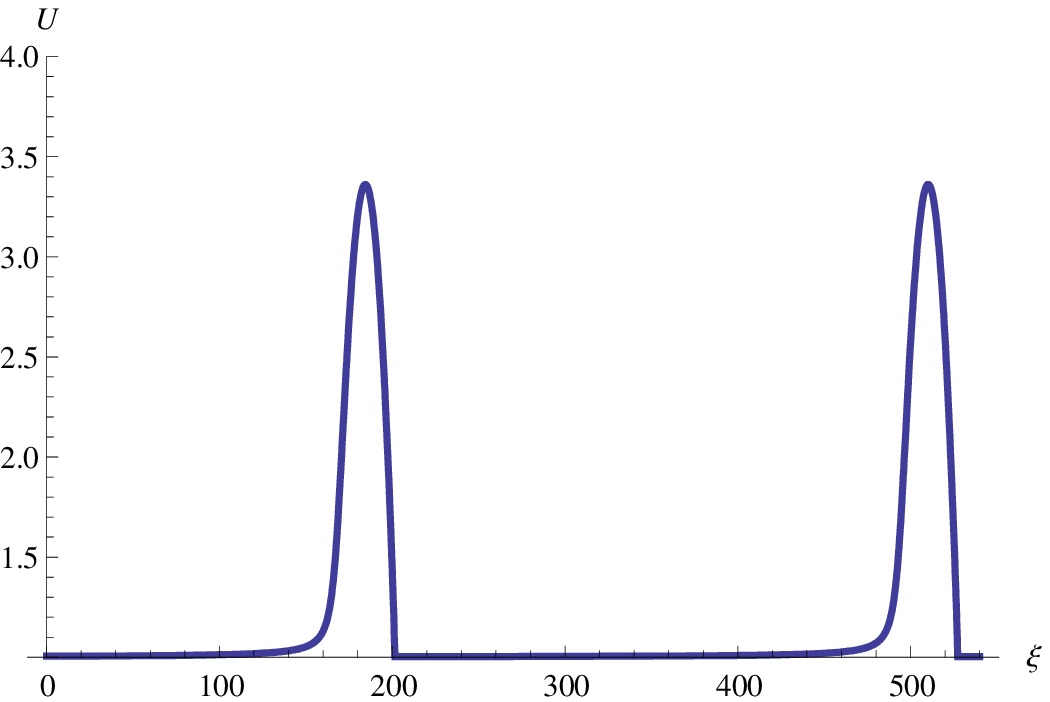}
\includegraphics[totalheight=1. in,origin=c]{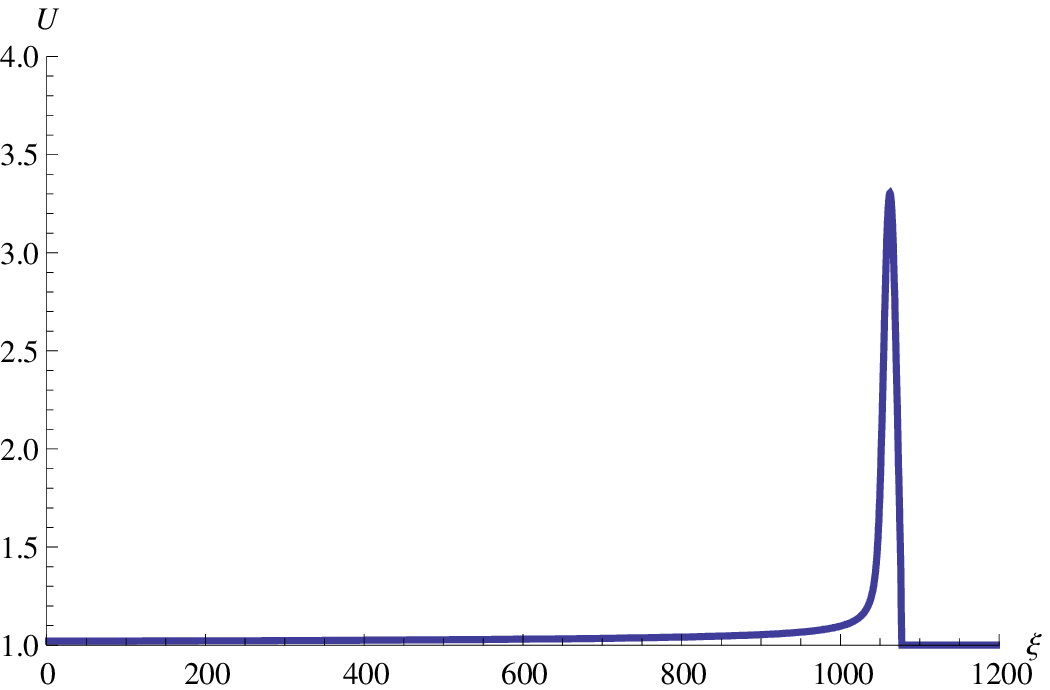}

\caption{Shock-like solutions to Eq. (\ref{GBE1}), corresponding to
$\varphi(U_0+X)=X^m$, $U_* \approx U_0$, $n=4$,  $m=1$ (left), $m=2$
(center), and $m=3$ (right)}\label{fig:8}

\end{center}
\end{figure}

Fig.~~\ref{fig:8} shows the series obtained for $n=4$ and $m$
varying from 1 to 3. It is seen that the waves become more and more
localized as the parameter $m$ grows. This series mainly
characterizes the growing stability of the saddle point $(0,\,0)$.

\begin{figure}
\begin{center}
\includegraphics[totalheight=1.5 in]{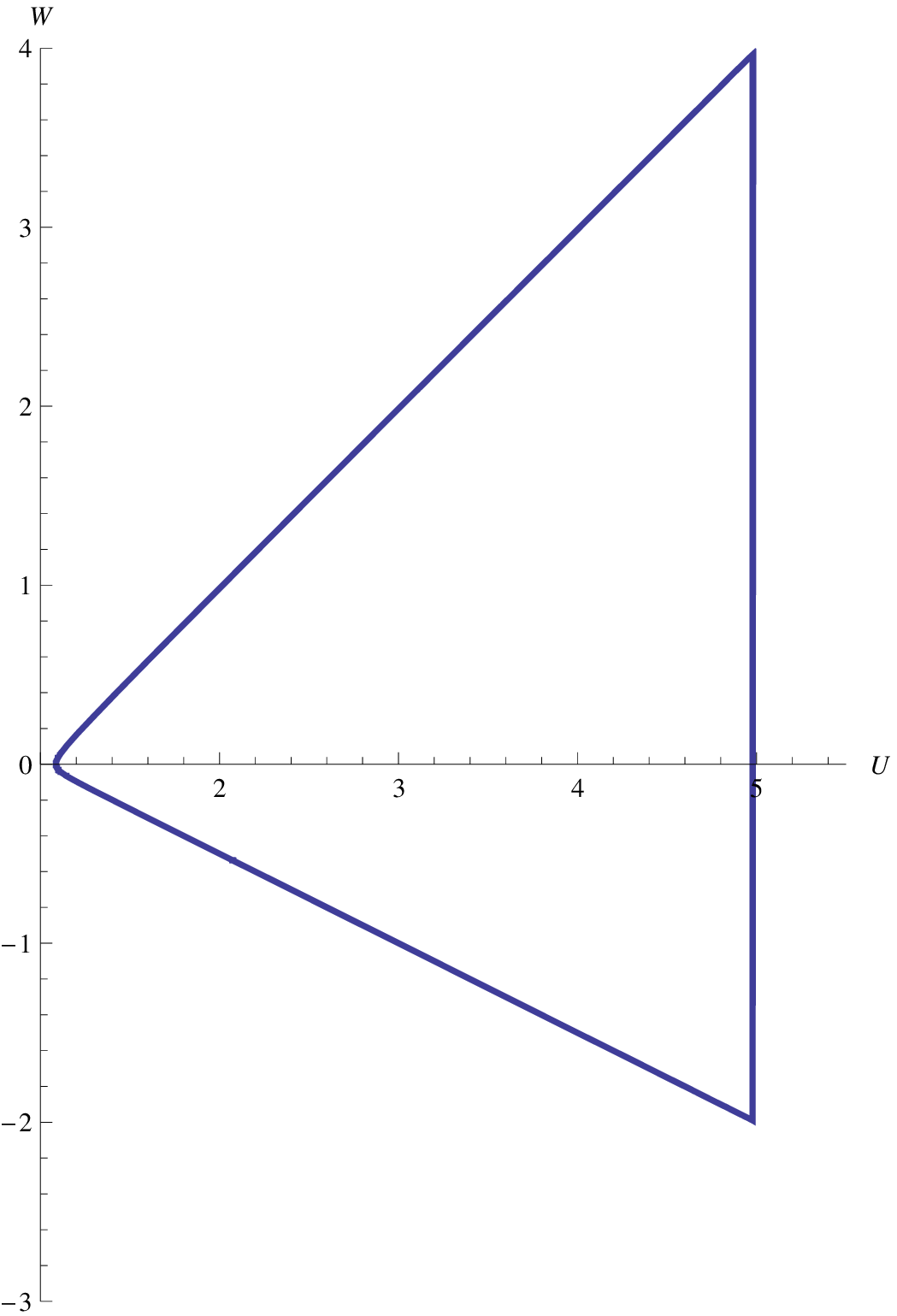}
\hspace{5 mm}
\includegraphics[totalheight=1.5 in,origin=c]{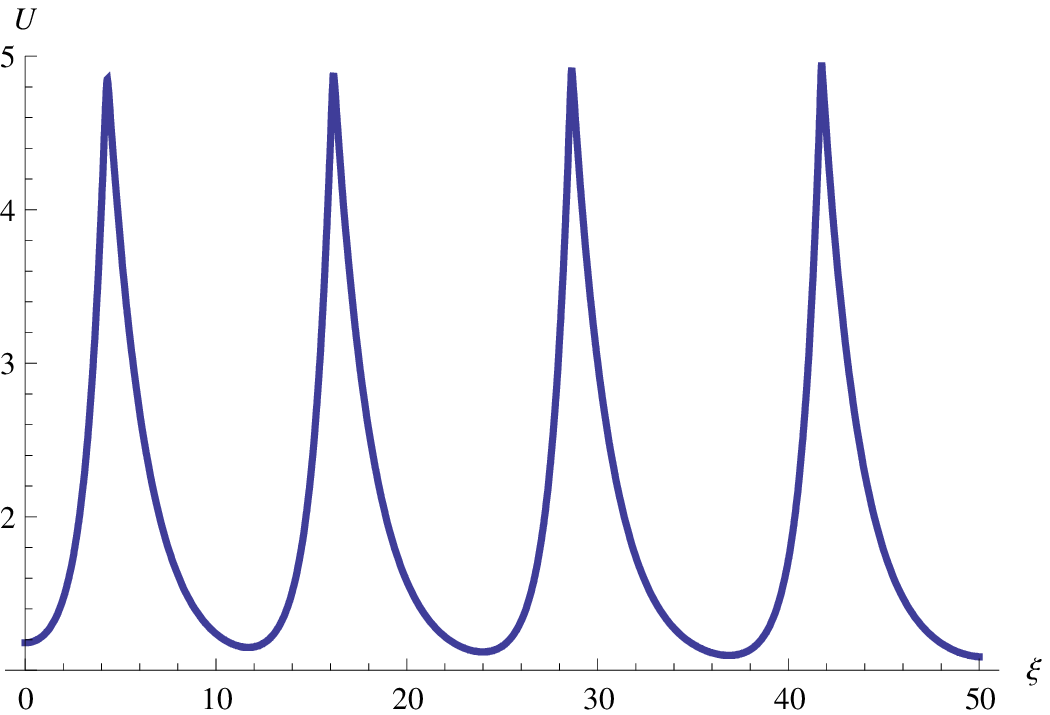}
\caption{Periodic solution of the system (\ref{factors2}) with
$\varphi(U_0+X)=-X^1$ (left) and the corresponding tandem of
generalized cusp-like solutions to Eq. (\ref{GBE1}) (right),
obtained for $n=1$, $\alpha=0.552$, $V_{cr_2} \cong 3.00593$ and
$U_*-U_1=1.98765$ }\label{fig:9}

\end{center}
\end{figure}

\begin{figure}
\begin{center}
\includegraphics[totalheight=1.5 in]{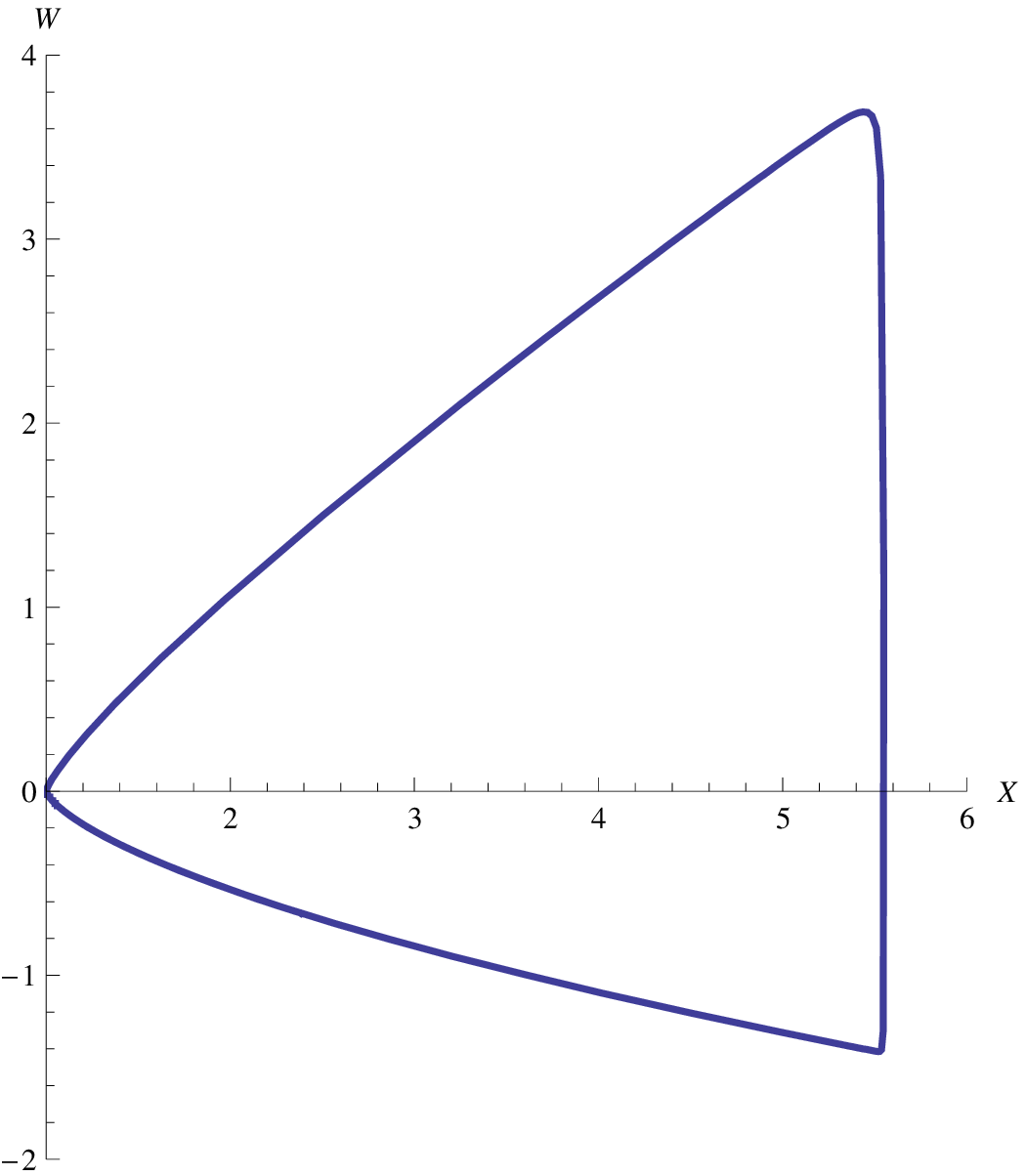}
\hspace{5 mm}
\includegraphics[totalheight=1.5 in,origin=c]{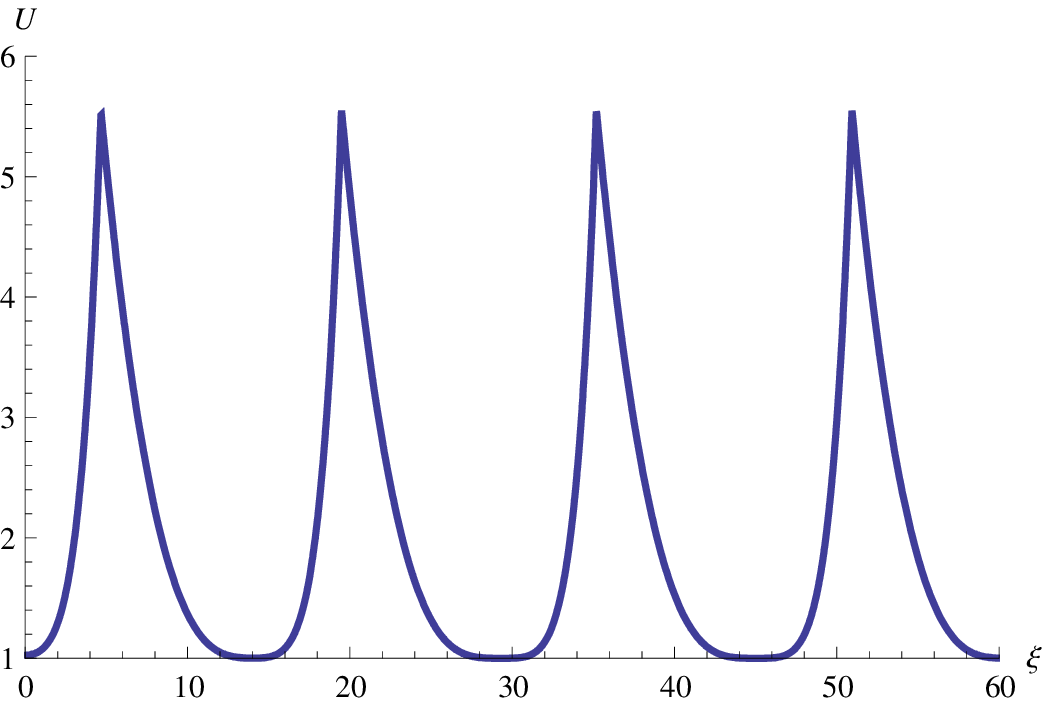}
\caption{Homoclinic solution of the system (\ref{factors2}) with
$\varphi(U_0+X)=-X^{\frac{1}{2}}$ (left) and the corresponding
tandem of generalized cusp-like solutions to Eq. (\ref{GBE1})
(right), obtained for $n=1$, $\alpha=0.562$, $V_{cr_2} \cong
3.14497$ and $U_*-U_1=2.55863$ }\label{fig:10}

\end{center}
\end{figure}

Finally, let us discuss what happens when
$U_*\left(V_{cr_1}\right)>U_1$. The birth of the stable limit cycle
is possible in this case, if $\varphi(U)<0$. Therefore we use in
numerical experiment the function $\varphi(U_0+X)=-X^m$. Numerical
study shows, that the radius of the limit cycle grows as the
bifurcation parameter $V$ grows. Simultaneously the coordinate
$U_*(V)$ of the singular line intersection with the horizontal axis
moves to the right, but not so quickly as the radius of the limit
cycle. In most cases the destruction of the periodic trajectory
observed was due to its interaction with the line $\Delta(U)=0$. For
$m=n=1$ we succeeded in observing how the limit cycle approaches the
line of singularity, attaining the  triangle shape
Fig.~~\ref{fig:9}, left. The corresponding  succession of the
cusp-like travelling waves is shown in the right picture. Let us
note,  that we were not able to follow the moment of the homoclinic
bifurcation,  varying the  parameters  $\alpha$ and $V$.

Experimenting with $m=1/2,\,\,n=1$ occurs to be more easy. The
homoclinic bifurcation, shown in the left picture of the
Fig.\ref{fig:9},  takes place when the far end of the cycle attains
the line $\Delta(U)=0$. The presence of the singular line causes a
drastic change of the shape of  homoclinic loop, which equally well
can be called the "homoclinic triangle". The right picture shows the
corresponding succession of the well-separated peakons.

Let us note in conclusion that numerical  experiments performed for
the parabolic case, i.e. for $\alpha=0$, demonstrate quite similar
behavior of the TW and their dependence upon the parameters (for
more details, see \cite{ROMP_09}).

\section{Final remarks}

So it was shown in this study, that the generalized
convection-reaction-diffusion equation (\ref{GBE1}) possesses a wide
variety of TW solution, such as solitons, compactons, shock-like
solutions and peakons. The result of the qualitative analysis and
numerical simulation are in agreement with each other. The
hyperbolic case  occurs to be more reach, since the soliton-like
solutions and peakons are observed only when $\alpha >0$.
Surprisingly enough, one-sided compactons exist not only when the
stationary point lies on the singular line $\Delta(U)=0$, but also
when it lies left-hand side of it. Condition $m>1$ assures in the
last case the existence of the sharp front.

The results obtained are not rigorous. Only heuristic arguments are
attached in favor of the statement that for $m\neq 1/2$ an infinite
"time" is needed to reach the point $(U_0,\,\,0)$, moving along the
incoming separatrice of the saddle sector. The precision of the
numerical method used does not enable us to get convinced that we
deal with shock fronts with exponentially localized "tails" in cases
of small $m\in N_+$, when the critical point $(U_0,\,\,0)$
demonstrates essential instability and  the "trains" of sharply
ended pulses are observed, instead of truly localized solitary
waves. More precise description of the wave patterns seems to be
possible by presenting them in the form of exponential series
\cite{vladki05}. Another possibility is connected with the
application of the rigorous computing methods \cite{zglicz}. Aside
of the scope of this study remained the study of stability and
attracting properties of the invariant TW solutions. We plan to
address these issues in the forthcoming studies.


\end{document}